\documentclass[%
 reprint,
 amsmath,amssymb,
 aps,
pra,
]{revtex4-1}

\usepackage{graphicx}
\usepackage{dcolumn}
\usepackage{bm}

\usepackage{braket} 

\newcommand{\Ortho}{\mathrm{O}}
\newcommand{\SO}{\mathrm{SO}}
\newcommand{\U}{\mathrm{U}}
\newcommand{\SU}{\mathrm{SU}}

\begin{document}

\title{Semiclassical Analysis of the Wigner $12j$ Symbol \\ with One Small Angular Momentum}

\author{Liang Yu}%
\email{liangyu@wigner.berkeley.edu}
\affiliation{%
 Department of Physics, University of California, Berkeley, California 94720 USA \\
}%

\date{\today}

\begin{abstract}
We derive an asymptotic formula for the Wigner $12j$ symbol, in the limit of one small and 11 large angular momenta. There are two kinds of asymptotic formulas for the $12j$ symbol with one small angular momentum. We present the first kind of formula in this paper. Our derivation relies on the techniques developed in the semiclassical analysis of the Wigner $9j$ symbol [L.\ Yu and R.\ G.\ Littlejohn, Phys.\ Rev.\ A {\bf 83}, 052114 (2011)], where we used a gauge-invariant form of the multicomponent WKB wave-functions to derive asymptotic formulas for the $9j$ symbol with small and large angular momenta. When applying the same technique to the $12j$ symbol in this paper, we find that the spinor is diagonalized in the direction of an intermediate angular momentum. In addition, we find that the geometry of the derived asymptotic formula for the $12j$ symbol is expressed in terms of the vector diagram for a $9j$ symbol. This illustrates a general geometric connection between asymptotic limits of the various $3nj$ symbols. This work contributes the first known asymptotic formula for the $12j$ symbol to the quantum theory of angular momentum, and serves as a basis for finding asymptotic formulas for the Wigner $15j$ symbol with two small angular momenta.  
\begin{description}
\item[PACS numbers] 03.65.Sq, 02.30.Ik, 03.65.Vf
\end{description}
\end{abstract}

\pacs{03.65.Sq, 02.30.Ik, 03.65.Vf}
                             
\maketitle

\section{Introduction}

This paper derives in full detail one of two kinds of asymptotic formulas for the $12j$ symbol, in the limit of one small and 11 large angular momenta. We will first briefly review some of the previous works on the Wigner $12j$ symbol. Its definition and exact formula are described in the textbooks on angular momentum theory \cite{edmonds1960, brink1968, biedenharn1981, yutsis1962}. Although it is used less often than the Wigner $3j$-, $6j$ symbols, it has applications in the theory of x-ray absorption branching ratios \cite{thole1988}, two-photons absorption spectroscopy \cite{ceulemans1993}, and loop quantum gravity \cite{carfora1993}.  The $12j$ symbol was first defined in the paper by Jahn and Hope \cite{jahn1954} in 1954. In that paper, they listed two kinds of formulas for the special values of the $12j$ symbol when any one of its 12 arguments is zero. See Eq.\ (A8) and Eq.\ (A9) in \cite{jahn1954}. The complete symmetries of the $12j$ symbol are given in \cite{ord-smith1954}, as a result of the observation that the graphical representation of the triangular conditions of the $12j$ symbol is a M\"{o}bius strip. This M\"{o}bius strip is illustrated in Fig.\ \ref{ch7: fig_diagram_12j} below.

We note that Eq.\ (A8) in \cite{jahn1954} corresponds to placing the zero argument at the edge of this M\"{o}bius strip, and Eq.\ (A9) in \cite{jahn1954} corresponds to placing the zero argument at the center of the M\"{o}bius strip. Thus, we expect that, by placing the small angular momentum at the edge and center, respectively, of the M\"{o}bius strip, we will have two kinds of asymptotic formulas for the $12j$ symbol with one small angular momentum.

The main theoretical tool we use is a generalization of the Born-Oppenheimer approximation, in which the small angular momenta are the fast degrees of freedom and the large angular momenta are the slow degrees of freedom. The necessary generalization falls under the topic of multicomponent WKB theory. See \cite{littlejohn1991, littlejohn1992, weinstein1996} for the relevant background. The new techniques used in this paper are recently developed in the semiclassical analysis of the Wigner $9j$ symbol with small and large quantum numbers \cite{yu2011}. This paper makes extensive use of the results from that paper, and assumes a familiarity with it.

In analogy with the setup for the $9j$ symbol in \cite{yu2011}, we use exact linear algebra to represent the small angular momentum, and use the Schwinger's model to represent the large angular momenta. Each wave-function consists of a spinor factor and a factor in the form of a scalar WKB solution. For the $9j$ symbol, the scalar WKB solutions are represented by Lagrangian manifolds associated with a $6j$ symbol, which have been analyzed in \cite{littlejohn2010b} to reproduce the Ponzano Regge action \cite{ponzano-regge-1968}. In the problem of the $12j$ symbol with one small quantum number, the scalar WKB solutions are represented by Lagrangian manifolds associated with a $9j$ symbol. The actions for the $9j$ symbol are presented in \cite{littlejohn2010a}. We will quote their results in some of the semiclassical analysis of the $9j$ symbol in this paper.

We now give an outline of this paper. In section \ref{ch7: sec_12j_spin_network}, we display the spin network of the $12j$ symbol in the form of a M\"{o}bius strip, and decompose it into a scalar product of a bra and a ket. In section \ref{ch7: sec_12j_def}, we define the $12j$ symbol as a scalar product of two multicomponent wave-functions, whose WKB form are derived in section \ref{ch7: sec_12j_wave_fctn}. By following the procedure in \cite{yu2011}, we rewrite the multicomponent WKB wave-functions into their gauge-invariant forms in section \ref{ch7: sec_12j_gauge_inv_fctn}.  In section \ref{ch7: sec_lag_mfd_actions}, we describe the path used in a semiclassical analysis of the Lagrangian manifolds associated with the $9j$ symbol by generalizing the paths used in \cite{littlejohn2010b}. We then obtain the action integral associated with this path by quoting the results from \cite{littlejohn2010a}.  Finally, we calculate the spinor inner products at the intersections of the Lagrangian manifolds in section \ref{ch7: sec_spinor_products}. Putting the pieces together, we derive an asymptotic formula for the $12j$ symbol in section \ref{ch7: sec_12j_formula}, and display plots for this formula against exact values of the $12j$ symbol in section \ref{ch7: sec_plots}. The last section contains comments and discussions.

\section{\label{ch7: sec_12j_spin_network}Spin Network of the $12j$ Symbol}

\begin{figure}[tbhp]
\begin{center}
\includegraphics[width=0.33\textwidth]{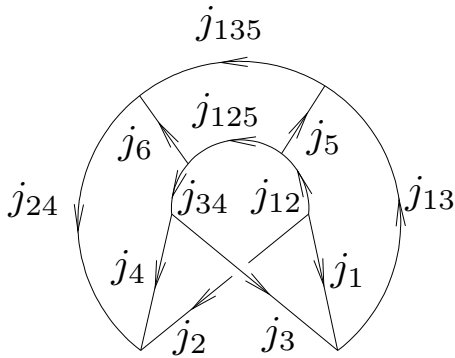}
\caption{The spin network of the Wigner $12j$ symbol.}
\label{ch7: fig_diagram_12j}
\end{center}
\end{figure}

The spin network \cite{yutsis1962}  for the Wigner $12j$ symbol 

\begin{equation}
\label{ch7: eq_12j_labeling}
\left\{
   \begin{array}{cccc}
    j_1 & j_2 & j_{12} & j_{125} \\ 
    j_3 & j_4 & j_{34} &  j_{135}  \\ 
    j_{13} & j_{24} & j_5 & j_6  \\
  \end{array} 
  \right \}    
\end{equation}
is illustrated in Fig.\ \ref{ch7: fig_diagram_12j}. In the spin network, each triangular condition of the $12j$ symbol is represented by a trivalent vertex. The spin network has the shape of a M\"{o}bius strip. 

The symmetries of the $12j$ symbol are associated with the symmetries of the M\"{o}bius strip, which are given by sliding along the M\"{o}bius strip and reflecting it about the vertical center of Fig.\ \ref{ch7: fig_diagram_12j}. Using these symmetries, any position in the center can be moved to any other position in the center, and any position on the edge can be moved to any other position on the edge. However, a center position cannot be move to an edge position, or vice versa. Thus, there are two inequivalent asymptotic limits of the $12j$ symbol with one small angular momentum, corresponding to placing the small angular momentum at the center or edge of the strip, respectively. In other words, we could either place the small angular momentum at $j_1, j_4, j_5$, or $j_6$ across the center of the strip, or place it at $j_2, j_3, j_{12}, j_{34}, j_{13}, j_{24}, j_{125}$, or $j_{135}$ along the edge of the strip. In this paper, we will focus on the case where the small angular momentum is placed at a center position at $j_5$.

\begin{figure}[tbhp]
\begin{center}
\includegraphics[width=0.33\textwidth]{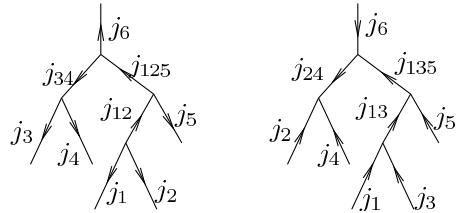}
\caption{A decomposition of the spin network of the $12j$ symbol.}
\label{ch7: fig_12j_decomposition}
\end{center}
\end{figure}

One can decompose the spin network of the $12j$ symbol into two spin network states by cutting $j_2, j_3$ at the twist, and cutting $j_1, j_4, j_5, j_6$ along the center of the strip. Using this decomposition, illustrated in Fig.\ \ref{ch7: fig_12j_decomposition}, the $12j$ symbol is expressed as a scalar product between a bra and a ket, in the Hilbert space represented by the six angular momenta $j_1, \dots, j_6$. This is explicitly expressed in Eq.\ (\ref{ch7: eq_12j_definition}).

\section{\label{ch7: sec_12j_def}Defining the $12j$ Symbol}

We use the decomposition of the spin network for the $12j$ symbol in Fig.\ \ref{ch7: fig_12j_decomposition} to write it as a scalar product. This is equivalent to Eq.\ (A4) in \cite{jahn1954}. We have

\begin{equation}
\label{ch7: eq_12j_definition}
\left\{
   \begin{array}{cccc}
    j_1 & j_2 & j_{12} & j_{125} \\ 
    j_3 & j_4 & j_{34} &  j_{135}  \\ 
    j_{13} & j_{24} & s_5 & j_6  \\
  \end{array} 
  \right \}    
= \frac{\braket{ b  |  a } }{ \{ [j_{12}][j_{34}][j_{13}][j_{24}] [j_{125}] [j_{135}] \}^{\frac{1}{2}}} \, , 
\end{equation}
where the square bracket notation $[\cdot]$ denotes $[c] = 2c + 1$, and $\ket{a}$ and $\ket{b}$ are normalized simultaneous eigenstates of lists of operators with certain eigenvalues. We will ignore the phase conventions of $\ket{a}$ and $\ket{b}$ for now, since we did not use them to derive our formula. In our notation, the two states are

\begin{equation}
\label{ch7: eq_a_state}
\ket{a} =  \left| 
\begin{array} { @{\,}c@{\,}c@{\,}c@{\,}c@{\;}c@{\;}c@{\;}c@{\;}c@{\;}c@{\;}c@{}}
	\hat{I}_1 & \hat{I}_2 & \hat{I}_3  & \hat{I}_4 &  {\bf S}_5^2 & \hat{I}_6 & \hat{\bf J}_{12}^2 & \hat{\bf J}_{34}^2 & \hat{\bf J}_{125}^2 &  \hat{\bf J}_{\text{tot}}  \\
	j_1 & j_2 & j_3 & j_4 & s_5 & j_6 & j_{12} & j_{34} & j_{125} & {\bf 0}  
\end{array}  \right>   \, , 
\end{equation}

\begin{equation}
\label{ch7: eq_b_state}
\ket{b} =  \left| 
\begin{array} { @{\,}c@{\,}c@{\,}c@{\,}c@{\;}c@{\;}c@{\;}c@{\;}c@{\;}c@{\;}c@{}}
	\hat{I}_1 & \hat{I}_2 & \hat{I}_3  & \hat{I}_4 &  {\bf S}_5^2 & \hat{I}_6 & \hat{\bf J}_{13}^2 & \hat{\bf J}_{24}^2 & \hat{\bf J}_{135}^2 &  \hat{\bf J}_{\text{tot}}  \\
	j_1 & j_2 & j_3 & j_4 & s_5 & j_6 & j_{13} & j_{24} & j_{135} & {\bf 0} 
\end{array}  \right>  \, .
\end{equation}
In the above notation, the large ket lists the operators on the top row, and the corresponding quantum numbers are listed on the bottom row. The hat is used to distinguish differential operators from their symbols, that is, the associated classical functions. 

The states $\ket{a}$ and $\ket{b}$ live in a total Hilbert space of six angular momenta ${\mathcal H}_1 \otimes {\mathcal H}_2 \otimes {\mathcal H}_3 \otimes {\mathcal H}_4 \otimes {\mathcal H}_6 \otimes {\mathcal H}_s$. Each large angular momentum ${\bf J}_r$, $r = 1,2,3,4,6$, is represented by a Schwinger Hilbert space of two harmonic oscillators, namely, ${\bf H}_r = L^2 ({\mathbb R}^2)$ \cite{littlejohn2007}. The small angular momentum ${\bf S}$ is represented by the usual $2s+1$ dimensional representation of $\SU(2)$, that is, ${\mathcal H}_s = {\mathbb C}^{2s+1}$, where $s = s_5$.

Let us now define the lists of operators in Eqs.\  (\ref{ch7: eq_a_state}) and Eq.\  (\ref{ch7: eq_b_state}). First we look at the operators $\hat{I}_r$, $r = 1,2,3,4,6$, ${\bf J}_{12}^2$, ${\bf J}_{34}^2$, ${\bf J}_{13}^2$, ${\bf J}_{24}^2$, which act only on the large angular momentum spaces ${\mathcal H}_r$, each of which can be viewed as a space of wave-functions $\psi(x_{r1}, x_{r2})$ for two harmonic oscillators of unit frequency and mass. Let $\hat{a}_{r\mu} = (\hat{x}_{r\mu} + i \hat{p}_{r\mu})/\sqrt{2}$, and $\hat{a}_{r\mu}^\dagger = (\hat{x}_{r\mu} - i \hat{p}_{r\mu})/\sqrt{2}$, $\mu = 1,2$, be the usual annihilation and creation operators. The operators $\hat{I}_r$ and $\hat{J}_{ri}$ are constructed from these differential operators $\hat{a}$ and $\hat{a}^\dagger$ as follows, 

\begin{equation}
\hat{I}_r = \frac{1}{2} \, \hat{a}_r^\dagger \hat{a}_r \, ,  \quad \quad \hat{J}_{ri} = \frac{1}{2} \, \hat{a}^\dagger_r \sigma_i \hat{a}_r  \, ,
\end{equation}
where $i = 1,2,3$, and $\sigma_i$ are the Pauli matrices. The quantum numbers $j_r$, $r = 1,2,3,4,6$ specify the eigenvalues of both $\hat{I}_r$ and $\hat{\bf J}_r^2$, to be $j_r$ and $j_r(j_r + 1)$, respectively.

The operators $\hat{\bf J}_{12}^2$, $\hat{\bf J}_{34}^2$, $\hat{\bf J}_{13}^2$, and $\hat{\bf J}_{24}^2$ that define intermediate coupling of the large angular momenta are defined by partial sums of $\hat{\bf J}_r$,

\begin{equation}
\label{ch7: eq_J12_J34_vector}
	\hat{\bf J}_{12} = \hat{\bf J}_1 + \hat{\bf J}_2 \, ,   \quad \quad  \hat{\bf J}_{34} = \hat{\bf J}_3 + \hat{\bf J}_4  \, .
\end{equation}

\begin{equation}
\label{ch7: eq_J13_J24_vector}
	\hat{\bf J}_{13} = \hat{\bf J}_1 + \hat{\bf J}_3 \, ,   \quad \quad  \hat{\bf J}_{24} = \hat{\bf J}_2 + \hat{\bf J}_4  \, .
\end{equation}

The quantum numbers $j_{i}$ , $i = 12, 34, 13, 24$ specify the eigenvalues of the operators $\hat{\bf J}_{i}^2$  to be $j_{i} (j_{i}+1)$, for $i = 12, 34, 13, 24$. See  \cite{littlejohn2007} for more detail on the Schwinger model.

Now we turn our attention to the operator $S^2$ that acts only on the small angular momentum space ${\mathbb C}^{2s+1}$. Let ${\bf S}$ be the vector of dimensionless spin operators represented by $2s+1$ dimensional matrices that satisfy the $\SU(2)$ commutation relations

\begin{equation}
[S_i, S_j] = i \, \epsilon_{ijk} \, S_k \, .
\end{equation}
The Casimir operator, ${\bf S}^2 = s (s+1)$, is proportional to the identity operator, so its eigenvalue equation is trivially satisfied.

The remaining operators ${\bf \hat{J}}_{125}^2$, ${\bf \hat{J}}_{135}^2$, and ${\bf \hat{J}}_{\text{tot}}$ are non-diagonal matrices of differential operators. They are defined in terms of the operators $\hat{I}_r$, $\hat{\bf J}_{ri}$, and ${\bf S}_i$ as follows,

\begin{eqnarray}
\label{ch7: eq_J125_square}
({\hat{J}}_{125}^2)_{\alpha \beta} &=& [  J_{12}^2+ \hbar^2 s(s+1) ] \delta_{\alpha \beta} + 2 {\bf \hat{J}}_{12} \cdot {\bf S}_{\alpha \beta} , \quad \quad  \\
\label{J135_sq}
({\hat{J}}_{135}^2)_{\alpha \beta}  &=& [  J_{13}^2 + \hbar^2 s(s+1) ] \delta_{\alpha \beta} + 2 {\bf \hat{J}}_{13} \cdot {\bf S}_{\alpha \beta} ,   \\
\label{ch7: eq_Jtot_vector}
({\bf \hat{J}}_{\text{tot}})_{\alpha \beta}  &=& ( {\bf \hat{J}}_1 + {\bf \hat{J}}_2 + {\bf \hat{J}}_3 + {\bf \hat{J}}_4 + {\bf \hat{J}}_6 ) \delta_{\alpha \beta} + \hbar \, {\bf S}_{\alpha \beta} .
\end{eqnarray}
These three operators act nontrivially on both the large and  small angular momentum Hilbert spaces.

\section{\label{ch7: sec_12j_wave_fctn}Multicomponent Wave-functions}

We follow the approach used in paper \cite{yu2011} to find a gauge-invariant form of the multicomponent wave-functions $\psi_{\alpha}^a (x) = \braket{x, \alpha | a}$ and $\psi_\alpha^b(x) = \braket{x, \alpha | b}$. Let us focus on $\psi_\alpha^a(x)$, since the treatment for $\psi^b$ is analogous. We will  drop the index $a$ for now.

Let $\hat{D}_i$, $i = 1, \dots, 12$ denote the the operators listed in the definition of $\ket{a}$ in Eq.\  (\ref{ch7: eq_a_state}). We seek a unitary operator $\hat{U}$, such that $\hat{D}_i$ for all $i=1, \dots, 12$ are diagonalized when conjugated by $\hat{U}$. In other words,   

\begin{equation}
\hat{U}^\dagger_{\alpha \, \mu} (\hat{D}_i)_{\alpha \, \beta} \, \hat{U}_{\beta \, \nu} = (\hat{\Lambda}_i)_{\mu \, \nu} \, , 
\end{equation}
where $\hat{\Lambda}_i$, $i =1, \dots, 12$ is a list of diagonal matrix operators. Let $\phi^{(\mu)}$ be the simultaneous eigenfunction for the $\mu^{\text{th}}$ diagonal entries $\hat{\lambda}_i$ of the operators $\hat{\Lambda}_i$, $i = 1, \dots, 12$. Then we obtain a simultaneous eigenfunction $\psi_\alpha^{(\mu)}$ of the original list of operators $\hat{D}_i$ from

\begin{equation}
\psi_\alpha^{(\mu)} = \hat{U}_{\alpha \, \mu} \, \phi^{(\mu)} \, . 
\end{equation}
Since we are interested in $\psi_\alpha$ only to first order in $\hbar$, all we need are the zeroth order Weyl symbol matrix $U$ of $\hat{U}$, and the first order symbol matrix $\Lambda_i$ of $\hat{\Lambda}_i$. The resulting asymptotic form of the wave-function $\psi(x)$ is a product of a scalar WKB part $B e^{iS}$ and a spinor part $\tau$, that is,

\begin{equation}
\label{ch7: eq_general_wave-function}
\psi_\alpha^{(\mu)} (x) =  B(x) \, e^{i \, S(x) / \hbar}  \, \tau_{\alpha}^{(\mu)} (x, p) \, . 
\end{equation}
Here the action $S(x)$ and the amplitude $B(x)$ are simultaneous solutions to the Hamilton-Jacobi and the transport equations, respectively, that are associated with the Hamiltonians $\lambda^{(\mu)}_i$. The spinor $\tau^{\mu}$ is the $\mu^{\text{th}}$ column of the matrix $U$, 

\begin{equation}
\label{ch7: eq_U_and_tau}
\tau_{\alpha}^{(\mu)} (x, p) = U_{\alpha \mu} (x, p) \, , 
\end{equation}
where $p = \partial S(x) / \partial x$. 

Now let us apply the above strategy to the $12j$ symbol. The Weyl symbols of the operators  $\hat{I}_r$ and $\hat{J}_{ri}$, $r = 1,2,3,4,6$, are $I_r - 1/2$ and $J_{ri}$, respectively, where 

\begin{equation}
\label{symbol_I_J}
I_r = \frac{1}{2} \,  \sum_\mu \overline{z}_{r\mu} z_{r\mu} \, ,   \quad \quad J_{ri} = \frac{1}{2} \, \sum_{\mu\nu} \overline{z}_{r\mu} (\sigma^i)_{\mu\nu} z_{r\nu} \, ,
\end{equation}
and where  $z_{r\mu} = x_{r\mu} + ip_{r\mu}$ and $\overline{z}_{r\mu} = x_{r\mu} - ip_{r\mu}$ are the symbols of $\hat{a}$ and $\hat{a}^\dagger$, respectively. The symbols of the remaining operators have the same expressions as Eqs.\  (\ref{ch7: eq_J12_J34_vector}), (\ref{ch7: eq_J13_J24_vector}), (\ref{ch7: eq_J125_square})-(\ref{ch7: eq_Jtot_vector}), but without the hats.

Among the operators $\hat{D}_i$, $\hat{J}_{125}^2$ and the vector of the three operators $\hat{\bf J}_{\rm tot}$ are non-diagonal. By looking at Eq.\ (\ref{ch7: eq_J125_square}), the expression for $\hat{J}_{125}^2$,  we see that the zeroth order term of the symbol matrix $J_{125}^2$ is already proportional to the identity matrix, so the spinor  $\tau$ must be an eigenvector for the first order term ${\bf J}_{12} \cdot {\bf S}$. Let $\tau^{(\mu)}({\bf J}_{12})$ be the eigenvector of the matrix ${\bf J}_{12} \cdot {\bf S}$ with eigenvalue $\mu J_{12}$, that is, it satisfies

\begin{equation}
\label{ch7: eq_JS_eigenvector}
({\bf J}_{12} \cdot {\bf S})_{\alpha \beta} \, \tau^{(\mu)}_\beta = \mu J_{12} \, \tau^{(\mu)}_\beta \, ,
\end{equation}
where $\mu = -s,\, \dots\, , \,+s$. In order to preserve the diagonal symbol matrices $J_{12}$ through the unitary transformation, we must choose the spinor $\tau^{(\mu)}$ to depend only on the direction of ${\bf J}_{12}$. One possible choice of $\tau^{(\mu)}$ is the north standard gauge, (see Appendix A of \cite{littlejohn1992}), in which the spinor $\delta_{\alpha\, \mu}$ is rotated along a great circle from the $z$-axis to the direction of ${\bf J}_{12}$. Explicitly, 

\begin{equation}
\label{tau_north_standard_gauge}
\tau^{(\mu)}_\alpha ({\bf J}_{12}) = e^{i (\mu - \alpha) \phi_{12}} \, d^{(s)}_{\alpha \, \mu} (\theta_{12}) \, , 
\end{equation}
where $(\theta_{12}, \phi_{12})$ are the spherical coordinates that specify the direction of ${\bf J}_{12}$.  Note that this is not the only choice, since Eq.\  (\ref{ch7: eq_JS_eigenvector}) is invariant under a local $\U(1)$ gauge transformations.  In other words, any other spinor $\tau' = e^{i g({\bf J}_{12})} \, \tau$ that is related to $\tau$ by a $\U(1)$ gauge transformation satisfies Eq.\  (\ref{ch7: eq_JS_eigenvector}).  This local gauge freedom is parametrized by the vector potential,

\begin{equation}
\label{def_A_vec}
{\bf A}^{(\mu)}_{12} = i (\tau^{(\mu)} )^\dagger \, \frac{\partial \tau^{(\mu)}}{\partial {\bf J}_{12}} \, , 
\end{equation}
which transforms as ${\bf A}^{(\mu)'} = {\bf A}^{(\mu)} - \nabla_{{\bf J}_{12}} (g)$ under a local gauge transformation. Moreover, the gradient of the spinor can be expressed in terms of the vector potential, (Eq.\ (A.22) in \cite{littlejohn1992}), as follows, 

\begin{equation}
\label{ch7: eq_tau_derivative}
\frac{\partial \tau^{(\mu)}}{\partial {\bf J}_{12}} = i \left( - {\bf A}_{12}^{(\mu)} + \frac{{\bf J}_{12} \times {\bf S}}{J_{12}^2} \right) \, \tau^{(\mu)} \, .
\end{equation}
Once we obtain the complete set of spinors $\tau^{(\mu)}$, $\mu = -s, \dots, s$, we can construct the zeroth order symbol matrix $U$ of the unitary transformation $\hat{U}$ from Eq.\  (\ref{ch7: eq_U_and_tau}). 

Now let us show that all the transformed symbol matrices of the operators in Eq.\  (\ref{ch7: eq_a_state}),  namely, the $\Lambda_i$, are diagonal to first order. Let us write $\hat{\Lambda} [ \hat{D} ]$ to denote the operator $\hat{U}^\dagger \hat{D} \hat{U}$, and write $\Lambda [ \hat{D} ]$ for its Weyl symbol. First, consider the operators $\hat{I}_r$, $r = 1,2,3, 4, 6$, which are proportional to the identity matrix. Using the operator identity

\begin{equation}
[\hat{\Lambda}(\hat{I}_r)]_{\mu\nu} = \hat{U}^\dagger_{\alpha \mu} ( \hat{I}_r  \delta_{\alpha \beta} ) \hat{U}_{\beta \nu} = \hat{I}_r \delta_{\mu \nu} - \hat{U}^\dagger_{\alpha \mu} [ \hat{U}_{\alpha \nu} , \, \hat{I}_r] \, ,
\end{equation}
we find

\begin{equation}
\label{ch7: eq_symbol_trick}
[\Lambda(\hat{I}_r)]_{\mu\nu}  = (I_r - 1/2) \delta_{\mu \nu} - i \hbar  U_{0\alpha \mu}^* \, \{ U_{0 \alpha \nu} , \, I_r \} \, , 
\end{equation}
where we have used the fact that the symbol of a commutator is a Poisson bracket. Since $U_{\alpha \mu} = \tau^{(\mu)}_\alpha$ is a function only of ${\bf J}_{12}$, and since the Poisson brackets $\{ {\bf J}_{12}, I_r \} = 0$ vanish for all $r = 1,2,3, 4, 6$, the second term in Eq.\  (\ref{ch7: eq_symbol_trick}) vanishes. We have

\begin{equation}
[\Lambda(\hat{I}_r)]_{\mu\nu} = (I_r - 1/2) \, \delta_{\mu\nu} \, .
\end{equation}
Similarly, because $\{ {\bf J}_{12}, J_{12}^2 \} = 0$ and $\{ {\bf J}_{12}, J_{34}^2 \} = 0$, we find

\begin{equation}
[\Lambda(\hat{J}_{12}^2 ) ]_{\mu\nu} = J_{12}^2 \, \delta_{\mu\nu} \, ,  \quad \quad
[\Lambda(\hat{J}_{34}^2 ) ]_{\mu\nu} = J_{34}^2 \, \delta_{\mu\nu} \, .
\end{equation}

Now we find the symbol matrices $\Lambda({\bf \hat{J}}_{125})$ for the vector of operators ${\bf \hat{J}}_{125}$, where 

\begin{equation}
[\hat{\Lambda}({\bf \hat{J}}_{125})]_{\mu \nu} = \hat{U}^\dagger_{\alpha \mu} ( {\bf \hat{J}}_{12} \delta_{\alpha \beta}) \hat{U}_{\beta \nu} + \hbar \, \hat{U}^\dagger_{\alpha \mu} {\bf S}_{\alpha \beta} \hat{U}_{\beta \nu} \, . 
\end{equation}
After converting the above operator equation to Weyl symbols, we find 

\begin{eqnarray}
\label{ch7: eq_J_125_Lambda}
&& [\Lambda({\bf \hat{J}}_{125})]_{\mu\nu}    \\  \nonumber
&=& {\bf J}_{12} \delta_{\mu \nu} - i \hbar U_{\alpha \mu}^* \{ U_{\alpha\mu } , \, {\bf J}_{12} \}  + \hbar \, U_{\alpha \mu}^* {\bf S}_{\alpha \beta}  U_{\beta \nu} \\  \nonumber
&=& {\bf J}_{12} \delta_{\mu \nu} - i \hbar \tau^{(\mu)*}_\alpha  \{ \tau^{(\nu)}_\alpha , \, {\bf J}_{12} \}  + \hbar \, \tau^{(\mu)*}_\alpha {\bf S}_{\alpha \beta}  \tau^{(\nu)}_\beta \, . 
\end{eqnarray}
Let us denote the second term above by $T^i_{\mu \nu}$, and use Eq.\ (\ref{ch7: eq_tau_derivative}), the orthogonality of $\tau$, 
\begin{equation}
\tau^{(\mu)*}_\alpha  \, \tau^{(\nu)}_\alpha = \delta_{\mu \nu} \, ,
\end{equation}
to get

\begin{eqnarray}
\label{ch7: eq_T_i}
T^i_{\mu \nu} &=&  - i \hbar \tau^{(\mu)*}_\alpha \{ \tau^{(\nu)}_\alpha  , \, J_{12i} \}  \\  \nonumber
&=& - i \hbar \tau^{(\mu)*}_\alpha [ \{ \tau^{(\nu)}_\alpha  , \, J_{1i} \} + \{ \tau^{(\nu)}_\alpha  , \, J_{2i} \} ] \\  \nonumber
&=&  - i  \hbar \tau^{(\mu)*}_\alpha   \epsilon_{kji}   \left(   J_{1k} \frac{\partial \tau^{(\nu)}_\alpha}{ \partial J_{1j} } +   J_{2k} \frac{\partial \tau^{(\nu)}_\alpha}{ \partial J_{2j} }  \right)  \\  \nonumber
&=&  - i  \hbar \tau^{(\mu)*}_\alpha   \epsilon_{kji}    J_{12k}  \frac{\partial \tau^{(\nu)}_\alpha}{ \partial J_{12j} }    \\   \nonumber
&=& \hbar ({\bf A}_{12}^{(\mu)}  \times {\bf J}_{12})_i \, \delta_{\mu \nu} + \hbar \frac{\mu {J}_{12i}}{J_{12}} \delta_{\mu \nu} - \hbar \, \tau^{(\mu)*}_\alpha {S}_{\alpha \beta}  \tau^{(\nu)}_\beta \, ,
\end{eqnarray}
where in the third equality, we have used the reduced Lie-Poisson bracket (Eq.\ (30) in \cite{littlejohn2007}) to evaluate the Poisson bracket $\{ \tau, {\bf J}_1 \}$ and $\{ \tau, {\bf J}_2 \}$, and in the third equality we used $\partial \tau / J_1 = \partial \tau / J_{12}$ and  $\partial \tau / J_1 = \partial \tau / J_{12}$ from the chain rule, and in the fifth equality, we used Eq.\  (\ref{ch7: eq_tau_derivative}) for $\partial \tau / \partial {\bf J}_{12}$. Notice the term involving ${\bf S}$ in $T^i_{\mu\nu}$ in Eq.\  (\ref{ch7: eq_T_i}) cancels out the same term in $\Lambda({\bf \hat{J}}_{125})$ in Eq.\  (\ref{ch7: eq_J_125_Lambda}), leaving us with a diagonal symbol matrix 

\begin{equation}
\label{ch7: eq_J125_vector}
[\Lambda({\bf \hat{J}}_{125})]_{\mu\nu} = {\bf J}_{12} \left[ 1 + \frac{\mu \hbar}{J_{12}} \right] + \hbar \, {\bf A}_{12}^{(\mu)} \times {\bf J}_{12} \, . 
\end{equation}
Taking the square, we obtain 

\begin{equation}
[ \Lambda({\bf \hat{J}}_{125}^2) ]_{\mu\nu}
=  ( J_{12}  + \mu \hbar)^2  \delta_{\mu \nu} \, . 
\end{equation}
Finally, let us look at the last three remaining operators ${\bf \hat{J}}_{\text{tot}}$ in Eq.\  (\ref{ch7: eq_Jtot_vector}). Since each of the the symbols ${\bf J}_r$ for $r = 3,4,6$ defined in Eq.\ (\ref{symbol_I_J}) Poisson commutes with ${\bf J}_{12}$, that is, $\{ {\bf J}_{12}, {\bf J}_r \} = 0$, we find $\Lambda({\bf \hat{J}}_r) = {\bf J}_r - i \hbar U_0^\dagger \{ U_0({\bf J}_{12}), {\bf J}_r \} = {\bf J}_r$, for $r = 3, 4, 6$. Using $\Lambda({\bf \hat{J}}_{125})$ from Eq.\  (\ref{ch7: eq_J125_vector}), we obtain 
 
\begin{eqnarray}
&& [\Lambda({\bf \hat{J}}_{\text{tot}})]_{\mu\nu}     \\   \nonumber
&=& \left[  {\bf J}_{12}  \left( 1 + \frac{\mu \hbar}{J_{12}} \right) + \hbar \, {\bf A}_{12}^{(\mu)} \times {\bf J}_{12} + ({\bf J}_3 + {\bf J}_4 + {\bf J}_6 ) \right]  \delta_{\mu \nu} \, .
\end{eqnarray}
Therefore, all $\Lambda_i$, $i = 1,\dots, 12$, are diagonal. 

The analysis above is completely analogous to those in \cite{yu2011}, except that the spinor is diagonalized in the direction of the intermediate angular momentum vector ${\bf J}_{12}$. We see that the procedure in \cite{yu2011} generalizes to the case of the $12j$ symbol wave-functions without any complication. This is because of the chain rule for differentiation and Poisson brackets. See the calculations in Eq.\ (\ref{ch7: eq_T_i}). 

Not counting the trivial eigenvalue equation for $S^2$, we have $11$ Hamilton-Jacobi equations associated with the $\Lambda_i$ for each polarization $\mu$ in the $20$ dimensional phase space  ${\mathbb C}^{10}$. It turns out that not all of them are functionally independent. In particular, the Hamilton-Jacobi equations $\Lambda(\hat{J}_{12}^2) = J_{12}^2 \hbar = ( j_{12} + 1/2 ) \hbar$ and $\Lambda(\hat{J}_{125}^2) = (J_{12} + \mu \hbar)^2 = ( j_{125} + 1/2 )^2 \hbar^2$ are functionally dependent. For them to be consistent, we must pick out the polarization $\mu = j_{125} - j_{12}$. This reduces the number of independent Hamilton-Jacobi equations for $S(x)$ from $11$ to $10$, half of the dimension of the phase space ${\mathbb C}^{10}$. These ten equations define the Lagrangian manifold associated with the action $S(x)$. 

Now let us restore the index $a$. We express the multicomponent wave-function $\psi^a_\alpha(x)$ in the form of Eq.\  (\ref{ch7: eq_general_wave-function}), 

\begin{equation}
\label{ch7: eq_general_wave-function_a}
\psi^a_\alpha (x) = B_a(x) \, e^{i S_a(x) / \hbar} \,  \tau^a_\alpha(x, p)  \, . 
\end{equation}
Here the action $S_a(x)$ is the solution to the ten Hamilton-Jacobi equations associated with the $\mu^{\text{th}}$ entries $\lambda_i^a$ of ten of the symbol matrices $\Lambda_i^a$, given by
\begin{eqnarray}
\label{HJ_S_a}
I_1 &=& (j_1 + 1/2) \hbar \, ,  \\    \nonumber
I_2 &=& (j_2 + 1/2) \hbar \, ,  \\  \nonumber
I_3 &=& (j_3 + 1/2) \hbar \, ,  \\  \nonumber
I_4 &=& (j_4 + 1/2) \hbar \, ,  \\  \nonumber
I_6 &=& (j_6 + 1/2) \hbar \, ,  \\  \nonumber
J_{12}^2 &=& (j_{12} + 1/2)^2 \hbar^2 \, ,  \\  \nonumber
J_{34}^2 &=& (j_{34} + 1/2)^2 \hbar^2 \, ,  \\  \nonumber
{\bf J}_{\text{tot}}^{(a)} &=& {\bf J}_{12} \left[ 1 + \frac{\mu \hbar}{J_{12}} \right] + \hbar \, {\bf A}_{12} \times {\bf J}_{12} + ({\bf J}_3 + {\bf J}_4 + {\bf J}_6 ) = {\bf 0} \, ,
\end{eqnarray}
and $\tau^a = \tau^{(\mu)}$ with $\mu = j_{125}-j_{12}$. Note that all the Hamiltonians except the last three, ${\bf J}_{\text{tot}}^{(a)}$, preserve the value of ${\bf J}_{12}$ and ${\bf J}_6$ along their Hamiltonian flows.

We carry out an analogous analysis for $\psi^b(x)$. The result is
\begin{equation}
\label{ch7: eq_general_wave-function_b}
\psi^b_\alpha (x) = B_b(x) \, e^{i S_b(x) / \hbar} \,  \tau^b_\alpha(x, p) \, ,
\end{equation}
where $S_b(x)$ is the solution to the following $10$ Hamilton-Jacobi equations: 

\begin{eqnarray}
I_1 &=& (j_1 + 1/2) \hbar \, ,  \\    \nonumber
I_2 &=& (j_2 + 1/2) \hbar \, ,  \\  \nonumber
I_3 &=& (j_3 + 1/2) \hbar \, ,  \\  \nonumber
I_4 &=& (j_4 + 1/2) \hbar \, ,  \\  \nonumber
I_6 &=& (j_6 + 1/2) \hbar \, ,  \\  \nonumber
J_{13}^2 &=& (j_{13} + 1/2)^2 \hbar^2 \, ,  \\  \nonumber
J_{24}^2 &=& (j_{24} + 1/2)^2 \hbar^2 \, ,  \\  \nonumber
{\bf J}_{\text{tot}}^{(b)} &=& {\bf J}_{13} \left[ 1 + \frac{\nu \hbar}{J_{13}} \right] + \hbar \, {\bf A}_{13} \times {\bf J}_{13} + ({\bf J}_2 + {\bf J}_4 + {\bf J}_6 ) = {\bf 0} \, .
\end{eqnarray}
Here the spinor $\tau^b = \tau_b^{(\nu)}$ satisfies 

\begin{equation}
\label{ch7: eq_JS_eigenvector_b}
({\bf J}_{13} \cdot {\bf S})_{\alpha \beta} \, ( \tau^{(\nu)}_b)_\beta = \nu J_{13} \, ( \tau^{(\nu)}_b)_\beta \, ,
\end{equation}
where $\nu = j_{135}-j_{13}$.

The vector potential ${\bf A}_{13}$ is defined by

\begin{equation}
{\bf A}_{13} = i ( \tau^b )^\dagger \, \frac{\partial \tau^b }{\partial {\bf J}_{13}} \, .
\end{equation}
Again,  note that all the Hamiltonians except the last three, ${\bf J}_{\text{tot}}^{(b)}$, preserve the value of ${\bf J}_{13}$ and ${\bf J}_6$ along their Hamiltonian flows.

\section{\label{ch7: sec_12j_gauge_inv_fctn}The Gauge-Invariant Form of the Wave-functions}

We follow the procedure described by the analysis preceding Eq.\ (69) in \cite{yu2011} to transform the wave-functions into their gauge-invariant form. The result is a gauge-invariant representation of the wave-function,  

\begin{equation}
\label{wave_fctn_factorization_a}
\psi^a(x) = B_a(x) \, e^{i S_a^{9j}(x) / \hbar} \, \left[  U_a(x)  \, \tau^a(x_0) \right]   \, . 
\end{equation}
where the action $S_a^{9j}(x)$ is the integral of $p \, dx$ starting at a point $z_0$, which is the lift of a reference point $x_0$ in the Lagrangian manifold ${\mathcal L}_a^{9j}$. The Lagrangian manifold ${\mathcal L}_a^{9j}$ is defined by the following equations:

\begin{eqnarray}
\label{HJ_S_a_9j}
I_1 &=& (j_1 + 1/2) \hbar \, ,  \\    \nonumber
I_2 &=& (j_2 + 1/2) \hbar \, ,  \\  \nonumber
I_3 &=& (j_3 + 1/2) \hbar \, ,  \\  \nonumber
I_4 &=& (j_4 + 1/2) \hbar \, ,  \\  \nonumber
I_6 &=& (j_6 + 1/2) \hbar \, ,  \\  \nonumber
J_{12}^2 &=& (j_{12} + 1/2)^2 \hbar^2 \, ,  \\  \nonumber
J_{34}^2 &=& (j_{34} + 1/2)^2 \hbar^2 \, ,  \\  \nonumber
{\bf J}_{\text{tot}} &=& {\bf J}_1 + {\bf J}_2  + {\bf J}_3 + {\bf J}_4 + {\bf J}_6  = {\bf 0} \, .
\end{eqnarray}
The rotation matrix $U_a(x)$ that appears in Eq.\ (\ref{wave_fctn_factorization_a}) is determined by the $\SO(3)$ rotation that transforms the shape configuration of ${\bf J}_{12}$ and ${\bf J}_6$ at the reference point $z_0 = (x_0, p(x_0))$ on ${\mathcal L}_a^{9j}$ to the shape configuration of ${\bf J}_{12}$ and ${\bf J}_6$ at the point $z = (x, p(x))$ on ${\mathcal L}_a^{9j}$. Here ${\bf J}_{12}$ and ${\bf J}_6$ are functions of $z$ and are defined in Eq.\ (\ref{symbol_I_J}).

Similarly, the multicomponent wave-function for the state $\ket{b}$ has the following form,

\begin{equation}
\label{wave_fctn_factorization_b}
\psi^b(x) = B_b(x) \, e^{i S_b^{9j}(x) / \hbar} \, \left[ U_b(x)   \, \tau^b(x_0)  \right] \, , 
\end{equation}
where the action $S_b^{9j}(x)$ is the integral of $p \, dx$ starting at a point that is the lift of $x_0$ onto the Lagrangian manifold ${\mathcal L}_b^{9j}$. The Lagrangian manifold ${\mathcal L}_b^{9j}$ is defined by the following equations:

\begin{eqnarray}
\label{HJ_S_b_9j}
I_1 &=& (j_1 + 1/2) \hbar \, ,  \\    \nonumber
I_2 &=& (j_2 + 1/2) \hbar \, ,  \\  \nonumber
I_3 &=& (j_3 + 1/2) \hbar \, ,  \\  \nonumber
I_4 &=& (j_4 + 1/2) \hbar \, ,  \\  \nonumber
I_6 &=& (j_6 + 1/2) \hbar \, ,  \\  \nonumber
J_{13}^2 &=& (j_{13} + 1/2)^2 \hbar^2 \, ,  \\  \nonumber
J_{24}^2 &=& (j_{24} + 1/2)^2 \hbar^2 \, ,  \\  \nonumber
{\bf J}_{\text{tot}} &=& {\bf J}_1 + {\bf J}_2  + {\bf J}_3 + {\bf J}_4 + {\bf J}_6  = {\bf 0} \, .
\end{eqnarray}
The rotation matrix $U_b(x)$ that appears in Eq.\ (\ref{wave_fctn_factorization_b}) is determined by the $SO(3)$ rotation that transform the shape configuration of ${\bf J}_{13}$ and ${\bf J}_6$ at the reference point $z_0 = (x_0, p(x_0))$ on ${\mathcal L}_b^{9j}$ to the shape configuration of ${\bf J}_{13}$ and ${\bf J}_6$ at the point $z = (x, p(x))$ on ${\mathcal L}_b^{9j}$.

Taking the inner product of the wave-functions, and treating the spinors as part of the slowly varying amplitudes, we find 

\begin{eqnarray}
 \braket{b|a}  
&=&  e^{i\kappa} \sum_k    \Omega_k  \, \text{exp} \{i [S_a^{9j}(z_k) - S_b^{9j}(z_k) - \mu_k \pi /2 ] / \hbar \}   \nonumber  \\
&&  \left( U_b^{0k} \tau^b(z_0)\right)^\dagger  \left(U_a^{0k} \tau^a(z_0)\right)  .
\label{ch7: eq_general_formula}
\end{eqnarray}
In the above formula, the sum is over the components  of the intersection set ${\mathcal M}_k$ between the two Lagrangian manifolds ${\mathcal L}_a^{9j}$ and ${\mathcal L}_b^{9j}$. The point $z_k$ is any point in the $k$th component. The amplitude $\Omega_k$ and the Maslov index $\mu_k$ are the results of doing the stationary phase approximation of the inner product without the spinors. Each rotation matrix $U_a^{0k}$ is determined by a path $\gamma^{a (0k)}$ that goes from $z_0$ to $z_k$ along ${\mathcal L}_a^{9j}$, and $U_b^{0k}$ is similarly defined. The formula Eq.\  (\ref{ch7: eq_general_formula}) is independent of the choice of $z_k$, because any other choice $z_k'$ will multiply both $U_a^{0j}$ and $U_b^{0j}$ by the same additional rotation matrix which cancels out in the product $(U_b^{0k})^\dagger U_a^{0k}$.

The above analysis is a straightforward application of the theoretical result developed in \cite{yu2011}. We present the detail of this analysis to show that the procedure outlined in \cite{yu2011} does generalize to higher $3nj$ symbols, such as the $12j$ symbol.

\section{\label{ch7: sec_lag_mfd_actions}The Lagrangian Manifolds and Actions}

\begin{figure}[tbhp]
\begin{center}
\includegraphics[width=0.33\textwidth]{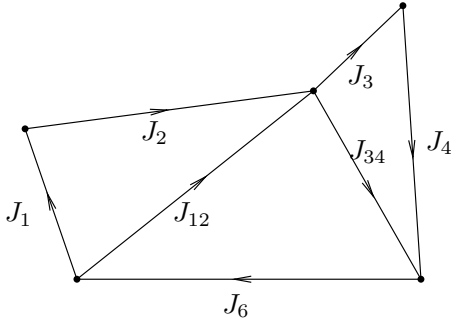}
\caption{The configuration of a point on ${\mathcal L}_a^{9j}$, projected onto the angular momentum space $\Lambda_{5j}$, and viewed in a single ${\mathbb R}^3$.}
\label{ch7: fig__9j_config_a}
\end{center}
\end{figure}

We now analyze the Lagrangian manifolds   ${\mathcal L}_a^{9j}$ and ${\mathcal L}_b^{9j}$, defined by the Hamilton-Jacobi equations Eq.\  (\ref{HJ_S_a_9j})  and  Eq.\  (\ref{HJ_S_b_9j}), respectively. We focus on  ${\mathcal L}_a^{9j}$ first, since the treatment for  ${\mathcal L}_b^{9j}$ is analogous. Let $\pi: \Phi_{5j} \rightarrow \Lambda_{5j}$ denote the projection of the large phase space  $\Phi_{5j} = ({\mathbb C}^2)^5$ onto the angular momentum space $\Lambda_{5j} = ({\mathbb R}^3)^5$, through the functions ${\bf J}_{ri}$, $r = 1,2,3,4,6$.  The first six equations, $I_r = j_r + 1/2$, $r = 1,2,3,4,6$ fix the lengths of the five vectors $|{\bf J}_r| = J_r$, $r = 1,2,3,4, 6$. The three equations for the total angular momentum, 

\begin{equation}
\label{ch7: eq_total_J_0}
{\bf J}_{\text{tot}} = {\bf J}_1 + {\bf J}_2  + {\bf J}_3 + {\bf J}_4 + {\bf J}_6  = {\bf 0} \, ,
\end{equation}
constrains the five vectors ${\bf J}_i$, $i = 1, \dots, 6$ to form a close polygon. The remaining two equations 

\begin{eqnarray}
J_{12}^2 &=& (j_{12} + 1/2)^2 \hbar^2 \, ,  \\ 
J_{34}^2 &=& (j_{34} + 1/2)^2 \hbar^2 \, ,  
\end{eqnarray}
put the vectors ${\bf J}_1, {\bf J}_2$ into a 1-2-12 triangle, and put the vectors ${\bf J}_3, {\bf J}_4$ into a 3-4-34 triangle. Thus, the vectors form a butterfly shape, illustrated in Fig.\ \ref{ch7: fig__9j_config_a}. This shape has two wings $(J_1,  J_2, J_{12})$ and $(J_3, J_4, J_{34})$ that are free to rotate about the $J_{12}$ and $J_{34}$ edges, respectively. Moreover, the Hamilton-Jacobi equations are also invariant under an overall rotation of the vectors. Thus the projection of ${\mathcal L}_a^{9j}$ onto the angular momentum space is diffeomorphic to $\U(1)^2 \times \Ortho(3)$.

The orbit of the group $\U(1)^5$ generated by $I_r$, $r = 1,2,3,4,6$ is a five-torus. Thus ${\mathcal L}_a^{9j}$ is a five-torus bundle over a sub-manifold described by the butterfly configuration in Fig.\ \ref{ch7: fig__9j_config_a}. Altogether there is a $\U(1)^7 \times \SU(2)$ action on ${\mathcal L}_a^{9j}$. If we denote coordinates on $\U(1)^7 \times \SU(2)$ by $(\psi_1, \psi_2, \psi_3, \psi_4, \psi_6, \theta_{12}, \theta_{34}, u)$, where $u \in \SU(2)$ and where the five angles are the $4\pi$-periodic evolution variables corresponding to $(I_1, I_2, I_3, I_4, I_6, {\bf J}_{12}^2, {\bf J}_{34}^2)$, respectively, then the isotropy subgroup is generated by three elements, say $x=(2 \pi, 2 \pi, 2 \pi, 2 \pi, 2 \pi, 0, 0, -1)$,  $y = (0, 0, 2 \pi, 2 \pi, 2 \pi, 2 \pi, 0, -1)$, and $z = (2 \pi, 2 \pi, 0, 0, 2 \pi, 0, 2 \pi, -1)$. The isotropy subgroup itself is an Abelian group of eight elements, $({\mathbb Z}_2)^3 = \{ e, x, y, z, xy, xz, yz, xyz \}$. Thus the manifold ${\mathcal L}_a^{9j}$  is topologically $\U(1)^7 \times \SU(2) / ({\mathbb Z}_2)^3$. The analysis for ${\mathcal L}_b^{9j}$ is the same. 

Now it is easy to find the invariant measure on ${\mathcal L}_a^{9j}$ and ${\mathcal L}_b^{9j}$. It is $d \psi_1 \wedge d \psi_2 \wedge d\psi_3 \wedge d\psi_4 \wedge d \psi_6 \wedge d\theta_{12} \wedge d \theta_{34} \wedge du$, where $du$ is the Haar measure on $\SU(2)$. The volumes $V_A$ of ${\mathcal L}_a^{9j}$ and $V_B$ of ${\mathcal L}_b^{9j}$ with respect to this measure are 

\begin{equation}
V_A = V_B = \frac{1}{8} \, (4 \pi)^7 \times 16 \pi^2 = 2^{15} \pi^9  \, , 
\end{equation}
where the $1/8$ factor compensates for the eight-element isotropy subgroup.

We now examine the intersections of ${\mathcal L}_a^{9j}$ and ${\mathcal L}_b^{9j}$ in detail. Because the two lists of Hamilton-Jacobi equations   (\ref{HJ_S_a_9j})  and (\ref{HJ_S_b_9j}) share the common equations $I_r = j_r + 1/2$, $r = 1,2,3,4,6$, the intersection in the large phase space $\Phi_{5j}$ is a five-torus fiber bundle over the intersection of the projections in the angular momentum space $\Lambda_{5j}$. The intersections of the projections in $\Lambda_{5j}$ require the five vectors ${\bf J}_r$, $r = 1,2,3,4,6$, to satisfy

\begin{eqnarray}
\label{ch7: eq_vector_equations}
| {\bf J}_r | = J_r \, , \quad\quad\quad  \sum_r  {\bf J}_r = {\bf 0}  \, ,  \\  \nonumber 
|{\bf J}_1 + {\bf J}_2 | = J_{12}  \, , \quad \quad  |{\bf J}_3 + {\bf J}_4 | = J_{34}  \, ,   \\  \nonumber
|{\bf J}_1 + {\bf J}_3 | = J_{13}  \, , \quad \quad  |{\bf J}_2 + {\bf J}_4 | = J_{24}   \, .
\end{eqnarray}
A nice way of constructing the vectors satisfying Eq.\ (\ref{ch7: eq_vector_equations}) follows the procedure given in the appendix of \cite{littlejohn2009}, which was generalized to apply to the symmetric treatment of the $9j$ symbol in \cite{littlejohn2010a}. For completeness, we summarize the construction in \cite{littlejohn2010a} using the unsymmetrical labeling of the $9j$ symbol in this paper in the next few paragraphs.

The construction uses the Gram matrix $G$ of dot products among the four vectors ${\bf J}_i$, $i = 1,2,3,4$. Some of the dot products are given by the length of the vectors $J_i$, $i = 1,2,3,4,6$, and the intermediate couplings $J_i$, $i = 12,34,13,24$. In particular, the diagonal elements are $J_i^2$, $i = 1,2,3,4$, and some of the off-diagonal elements are given by 

\begin{eqnarray}
{\bf J}_1 \cdot {\bf J}_2 &=& \frac{1}{2} ( J_{12}^2 - J_1^2 - J_2^2 ) \, , \\
{\bf J}_3 \cdot {\bf J}_4 &=& \frac{1}{2} ( J_{34}^2 - J_3^2 - J_4^2 ) \, ,   \\
{\bf J}_1 \cdot {\bf J}_3 &=& \frac{1}{2} (J_{13}^2 - J_1^2 - J_3^2)  \, , \\
 {\bf J}_2 \cdot {\bf J}_4 &=&  \frac{1}{2} (J_{24}^2 - J_2^2 - J_4^2 )  \, .
\end{eqnarray}
Let us denote the remaining two unknown dot products by $x = {\bf J}_1 \cdot {\bf J}_4$ and $y = {\bf J}_2 \cdot {\bf J}_3$. We have

\begin{widetext}
\begin{eqnarray}
 G 
 &=&  \left(
  \begin{array}{cccc}
    J_1^2 & \frac{1}{2} ( J_{12}^2 - J_1^2 - J_2^2 ) & \frac{1}{2} (J_{13}^2 - J_1^2 - J_3^2) &   x  \\ 
     \frac{1}{2} ( J_{12}^2 - J_1^2 - J_2^2 )  & J_2^2 & y & \frac{1}{2} (J_{24}^2 - J_2^2 - J_4^2 ) \\ 
    \frac{1}{2} (J_{13}^2 - J_1^2 - J_3^2)  & y  & J_3^2  & \frac{1}{2} ( J_{34}^2 - J_3^2 - J_4^2 )  \\
    x  & \frac{1}{2} (J_{24}^2 - J_2^2 - J_4^2 )  & \frac{1}{2} ( J_{34}^2 - J_3^2 - J_4^2 )  & J_4^2 \\
  \end{array} 
  \right) \, .   \label{ch7: eq_gram_matrix}
\end{eqnarray}
\end{widetext}
The unknown dot products $x$ and $y$ can be solved from a system of two equations. The first equation follows from Eq.\ (\ref{ch7: eq_total_J_0}). Moving ${\bf J}_6$ to the other side, and taking the square, it becomes

\begin{eqnarray}
J_6^2 &=& ({\bf J}_1 + {\bf J}_2 + {\bf J}_3 + {\bf J}_4)^2   \\  \nonumber
&=& J_1^2 + J_2^2 + J_3^2 + J_4^2 + 2 {\bf J}_1 \cdot {\bf J}_2 + 2 {\bf J}_3 \cdot {\bf J}_4   \\   \nonumber
&& \,  + 2  {\bf J}_1 \cdot {\bf J}_3 + 2 {\bf J}_2 \cdot {\bf J}_4 + 2x + 2y  \, , 
\end{eqnarray}
which gives us a linear relation between $x$ and $y$,

\begin{equation}
\label{ch7: eq_x_y_relation}
x + y = \frac{1}{2} (J_1^2 + J_2^2 + J_3^2 + J_4^2 + J_6^2 - J_{12}^2 - J_{34}^2 - J_{13}^2 - J_{24}^2) \, . 
\end{equation}
This is the same equation as Eq.\ (6) in \cite{littlejohn2010a}, except for the relabelling of the vectors. The second equation comes from the fact that the Gram matrix of the dot products between any four vectors in ${\mathbb R}^3$ has a zero determinant. That is, 

\begin{equation}
\label{ch7: eq_G_det_eq_0}
P(x, y)\equiv |G| = 0 \, . 
\end{equation}

This constitutes a second equation for $x$ and $y$. Substituting the linear relation Eq.\  (\ref{ch7: eq_x_y_relation}) into Eq.\  (\ref{ch7: eq_G_det_eq_0}) leads to a quartic equation $Q(x) = 0$, which we can use to solve for $x$. Then we can use Eq.\  (\ref{ch7: eq_x_y_relation}) to solve for $y$. In general, we find two sets of real solutions of $(x, y) = (x_1, y_1)$ and $(x, y) = (x_2, y_2)$. See \cite{littlejohn2010a} for more detail.

For each set of solutions of $(x, y)$, we obtain all the dot products among the first four vectors. Assuming all the diagonal sub-determinants of order $3$ of the Gram matrix in Eq.\  (\ref{ch7: eq_gram_matrix}) are positive definite, we can follow the procedure outlined in the appendix of \cite{littlejohn2009} to obtain the vectors. Let $G_3$ be the first diagonal $3 \times 3$ sub-matrix of $G$. We use its singular decomposition to determine the vectors ${\bf J}_1, {\bf J}_2, {\bf J}_3$. We can then find ${\bf J}_4$ from the known dot products between ${\bf J}_i$, $i=1,2,3$ and ${\bf J}_4$. Finally, we obtain ${\bf J}_6$ from

\begin{equation}
 {\bf J}_6 = - ({\bf J}_1 + {\bf J}_2 + {\bf J}_3 + {\bf J}_4 ) \, .
\end{equation}
Once we have ${\bf J}_i$, $i=1,2,3,4,6$, we add them up pairwise to find the intermediate vectors ${\bf J}_i$, $i = 12, 34, 13, 24$. This completes the construction of all nine vectors in ${\mathbb R}^3$.

\begin{figure}[tbhp]
\begin{center}
\includegraphics[width=0.33\textwidth]{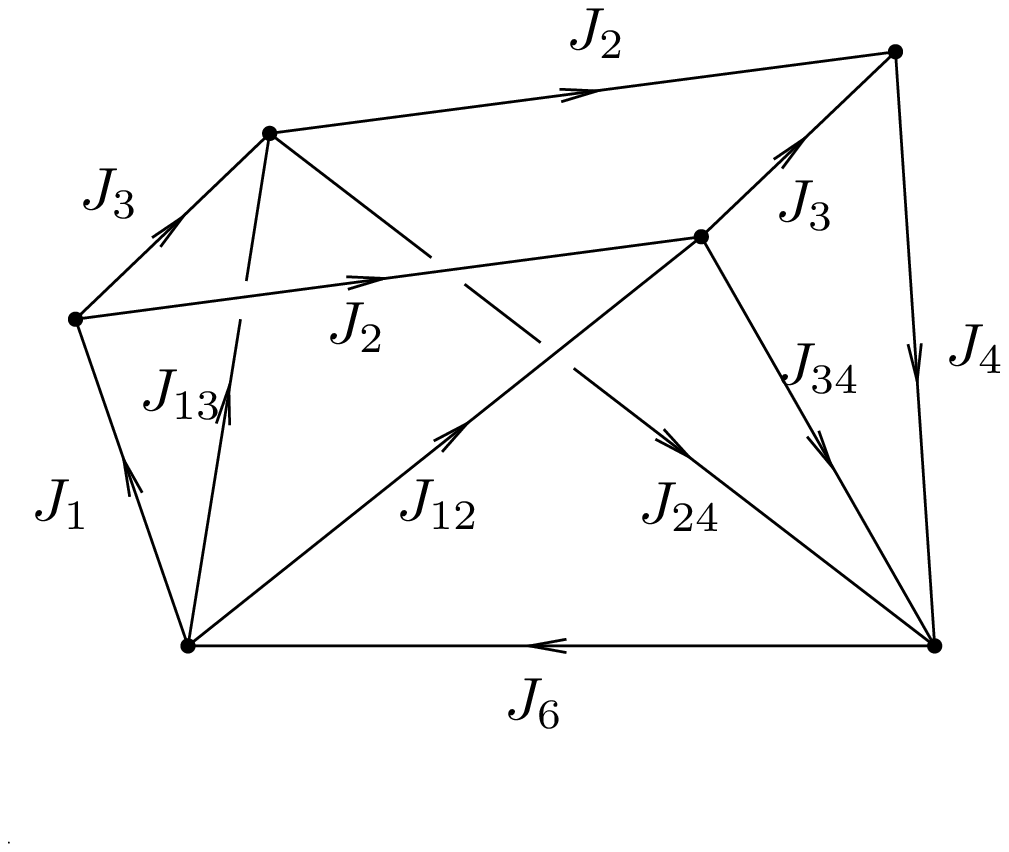}
\caption{The configuration of a point on the intersection $I_{11}$ set, projected onto the angular momentum space $\Lambda_{5j}$, and viewed in a single ${\mathbb R}^3$.}
\label{ch7: fig_9j_config_I_11}
\end{center}
\end{figure}

The construction of the vector configuration above not only gives explicit solutions for all the vectors at the intersection of ${\mathcal L}_a^{9j}$ and ${\mathcal L}_b^{9j}$, we also find that there are generally two distinct solutions of dot products $(x_1, y_1)$ and $(x_2, y_2)$. This implies that the solution set consists of two sets of vector configurations that are not related by an $O(3)$ symmetry. Thus the solution set of Eq.\  (\ref{ch7: eq_vector_equations}) in $\Lambda_{5j}$ consists of four disconnected subsets, each diffeomorphic to $SO(3)$. These four sets can be grouped into two pairs according to the values of the dot products $(x_1, y_1)$ and $(x_2, y_2)$.  

The intersections in $\Phi_{5j}$ are the lifts of the intersections in $\Lambda_{5j}$. Therefore, the intersection of ${\mathcal L}_a^{9j}$  consists of four disconnected subsets, where each subset is a five-torus bundle over $\SO(3)$. Let us denote the two sets corresponding to $(x_1, y_1)$ by $I_{11}, I_{12}$, and denote the two sets corresponding to $(x_2, y_2)$ by $I_{21}, I_{22}$. The vector configuration for a typical point in $I_{11}$ is illustrated in Fig.\ \ref{ch7: fig_9j_config_I_11}. Each intersection set is an orbit of the group $\U(1)^5 \times \SU(2)$, where $\U(1)^5$ represent the phases of the five spinors and $\SU(2)$ is the diagonal action generated by ${\bf J}_{\rm tot}$. 

The isotropy subgroup of this group action is ${\mathbb Z}_2$, generated by the element $ \; $ $(2 \pi, 2 \pi, 2\pi, 2\pi, 2\pi, -1)$, in coordinates $(\psi_1, \psi_2, \psi_3, \psi_4, \psi_6, u)$ for the group $\U(1)^5 \times \SU(2)$, where $u \in \SU(2)$. The volume of the intersection manifold $I_{11}$, $I_{12}$, $I_{21}$, or $I_{22}$, with respect to the measure $d \psi_1 \wedge d \psi_2 \wedge d\psi_3 \wedge d\psi_4 \wedge d \psi_6  \wedge du$, is 

\begin{equation}
V_I = \frac{1}{2} (4 \pi)^5 \times 16 \pi^2 = 2^{13} \pi^7   \, ,
\end{equation}
where the $1/2$ factor compensates for the two element isotropy subgroup. 

The amplitude determinant is given in terms of a determinant of Poisson brackets among distinct Hamiltonians between the two lists of Hamilton-Jacobi equations in Eqs.\  (\ref{HJ_S_a_9j})  and (\ref{HJ_S_b_9j}). In this case, those are $(J_{12}, J_{34})$ from Eq.\  (\ref{HJ_S_a_9j})  and $(J_{13}, J_{24})$ from  Eq.\  (\ref{HJ_S_b_9j}). Thus the determinant of Poisson brackets is
\begin{eqnarray}
&& \left|
  \begin{array}{cc}
    \{J_{12}, \, J_{13} \} & \{J_{12}, \, J_{24} \}   \\ 
    \{J_{34}, \, J_{13} \}  & \{J_{34}, \, J_{24} \}   \\ 
  \end{array} 
  \right|  \nonumber  \\   \nonumber
&=& \frac{1}{J_{12} J_{23} J_{13} J_{24} } \,
  \left|
  \begin{array}{cc}
    V_{123} & V_{214}  \\ 
    V_{341} & V_{432}   \\ 
  \end{array} 
  \right|  \\
  &=&  \frac{1}{J_{12} J_{23} J_{13} J_{24} }  | V_{123} V_{432} - V_{214} V_{341}  |  \, , 
\end{eqnarray}
where 

\begin{equation}
V_{ijk} = {\bf J}_i \cdot ({\bf J}_j \times {\bf J}_k)  \, . 
\end{equation}

The amplitude $\Omega_k$ in Eq.\  (\ref{ch7: eq_general_formula}) can be inferred from Eq.\ (10) in \cite{littlejohn2010a}. In the present case, each $\Omega_k$ has the same expression $\Omega$. It is 

\begin{eqnarray}
\Omega &=& \frac{(2 \pi i ) V_I}{\sqrt{V_A V_B} }  \, \frac{\sqrt{J_{12} J_{23} J_{13} J_{24}} }{\sqrt{| V_{123} V_{432} - V_{214} V_{341}  |} }  \nonumber   \\  \nonumber
&=& \frac{(2 \pi i ) 2^{13} \pi^7 }{ 2^{15} \pi^9 }  \, \frac{\sqrt{J_{12} J_{23} J_{13} J_{24}} }{\sqrt{| V_{123} V_{432} - V_{214} V_{341}  |} }     \\
&=& \frac{i \sqrt{J_{12} J_{23} J_{13} J_{24}}}{2 \pi \sqrt{| V_{123} V_{432} - V_{214} V_{341}  |} }  \, . 
\label{ch7: eq_amplitude}
\end{eqnarray}

We now outline the calculation of the relative phase between the exponents $S_a(z_{12}) - S_b(z_{12})$ and $S_a(z_{11}) - S_b(z_{11})$, which can be written as an action integral 

\begin{equation}
S^{(1)} = (S_a(z_{12}) - S_b(z_{12}) ) - (S_a(z_{11}) - S_b(z_{11}) )  = \oint \, p \, dx \, 
\end{equation}
around a closed loop that goes from $z_{11}$ to $z_{12}$ along ${\mathcal L}_a^{9j}$ and then back along ${\mathcal L}_b^{9j}$. 

\begin{figure}[tbhp]
\begin{center}
\includegraphics[width=0.45\textwidth]{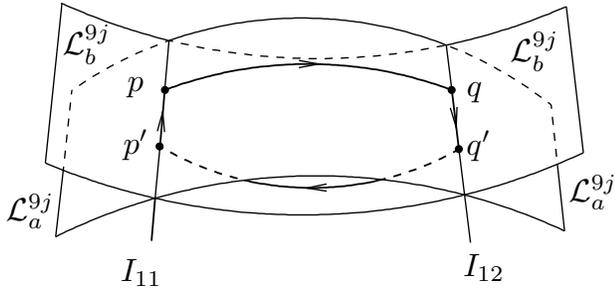}
\caption{The loop from a point $p \in I_{11}$ to $q \in I_{12}$ along ${\mathcal L}_a^{9j}$, and then to $q' \in I_{12}$ along $I_{12}$, and then to $p' \in I_{11}$ along ${\mathcal L}_b^{9j}$, and finally back to $p$ along $I_{11}$.}
\label{ch7: fig_loop_large_space}
\end{center}
\end{figure}

\begin{figure}[tbhp]
\begin{center}
\includegraphics[width=0.45\textwidth]{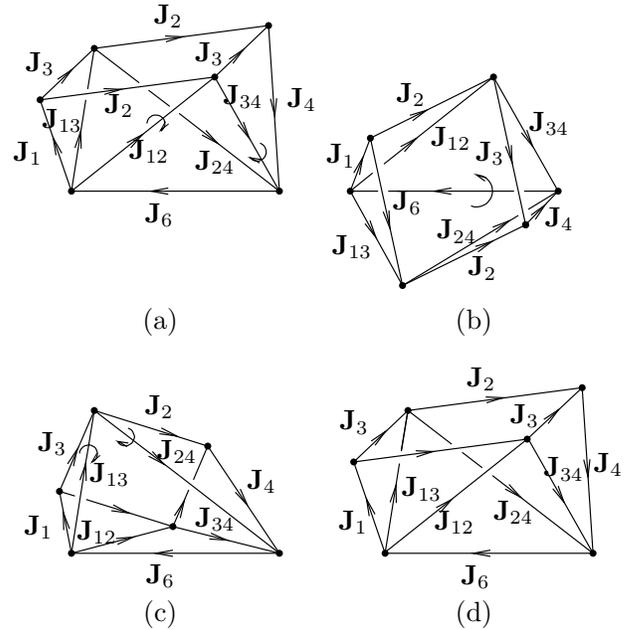}
\caption{The loop from Fig.\ \ref{ch7: fig_loop_large_space} projected onto a loop in $\Lambda_{5j}$, as viewed in a single ${\mathbb R}^3$.}
\label{ch7: fig_loop_small_space}
\end{center}
\end{figure}

We shall construct the closed loop giving the relative phase $S^{(1)}$ by following the Hamiltonian flows of various observables. This loop consists of four paths, and it is illustrated in the large phase space $\Phi_{5j}$ in Fig.\ \ref{ch7: fig_loop_large_space}. The loop projects onto a loop in the angular momentum space $\Lambda_{5j}$, which is illustrated in Fig.\ \ref{ch7: fig_loop_small_space}. We take the starting point $p \in I_{11}$ of Fig.\ \ref{ch7: fig_loop_large_space} to lie in the five-torus fiber above a solution of Eq.\  (\ref{ch7: eq_vector_equations}). Its vector configuration is illustrated in Fig.\ \ref{ch7: fig_loop_small_space}(a). 

First we follow the ${\bf J}_{12}^2$ flow and then the ${\bf J}_{34}^2$ flow to trace out a path that takes us along ${\mathcal L}_a^{9j}$ from a point $p$ in $I_{11}$ to a point $q$ in $I_{12}$. Let the angles of rotations be $2 \phi_{12}$ and $2 \phi_{34}$, respectively, where $\phi_{12}$ is the angle between the triangles 1-2-12 and 12-34-6, and $\phi_{34}$ is the angle between the triangles 3-4-34 and 12-34-6. These rotations effectively reflect the triangles 1-2-12 and 3-4-34  across the triangle 12-34-6, as illustrated in Figs.\ \ref{ch7: fig_loop_small_space}(a) and \ref{ch7: fig_loop_small_space}(b). In addition, the triangle 13-24-6 is also reflected across its own plane. Thus, all five vectors ${\bf J}_r$, $r = 1,2,3,4,6$, are reflected across the triangle 12-34-6.

Next, we follow the Hamiltonian flow generated by $-{\bf j}_6 \cdot {\bf J}_{\rm tot}$ along $I_{12}$, which generates an overall rotation of all the vectors around $- {\bf j}_6$. Let the angle of rotation be $2 \phi_6$, where $\phi_6$ is the angle between the triangles 12-34-6 and 13-24-6. This brings the triangle 13-24-6 back to its original position. However, the triangle 12-34-6 is now rotated to the other side of triangle 13-24-6, as illustrated in Fig.\  \ref{ch7: fig_loop_small_space}(c).  This corresponds to the point $q'$ in Fig. \ref{ch7: fig_loop_large_space}.

To bring the triangle 12-34-6 back to its original position, we follow the ${\bf J}_{13}^2$ flow and ${\bf J}_{24}^2$ flow along ${\mathcal L}_b^{9j}$. Let the angle of rotations be $2\phi_{13}$ and $2\phi_{24}$, respectively, where $\phi_{13}$ is the angle between the triangle 1-3-13 and the triangle 13-24-6, and $\phi_{24}$ is the angle between the triangle 2-4-24 and the triangle 13-24-6. These rotations effectively reflect all the vectors across the triangle 13-24-6. Thus we arrive at a point $p' \in I_{11}$, where the points $p$ and $p'$ have the same projection in the angular momentum space $\Lambda_{5j}$. This is illustrated in Figs.\ \ref{ch7: fig_loop_small_space}(a) and \ref{ch7: fig_loop_small_space}(d). Thus the two points $p$ and $p'$ differ only by the phases of the five spinors, which can be restored by following the Hamiltonian flows of $(I_1, I_2, I_3, I_4, I_6)$. This constitutes the last path from $p'$ to $p$.

To summarize the rotational history in the angular momentum space, we have applied the rotations

\begin{eqnarray}
\label{ch8: eq_rotations}
&& R_{13} ({\bf j}_{13}', 2 \phi_{13}) R_{24}({\bf j}_{24}', 2 \phi_{24})  R( - {\bf j}_6, 2 \phi_6) \\  \nonumber 
&& R_{34} ({\bf j}_{34}, 2 \phi_{34}) R_{12}({\bf j}_{12}, 2 \phi_{12})  \, , 
\end{eqnarray}
where $R_{12}$ acts only on ${\bf J}_1$ and ${\bf J}_2$, $R_{34}$ acts only on ${\bf J}_3$ and ${\bf J}_4$, $R_{13}$ acts only on ${\bf J}_1$ and ${\bf J}_3$, $R_{24}$ acts only on ${\bf J}_2$ and ${\bf J}_4$, and $R( - {\bf j}_6, 2 \phi_6 )$ acts on all five vectors. The corresponding $SU(2)$ rotations, with the same axes and angles, take us from point $p$ in Fig.\ \ref{ch7: fig_loop_large_space} to another point $p'$ along the sequence $p \rightarrow q \rightarrow q'  \rightarrow p'$.

To compute the final five phases required to close the loop, we use the Hamilton-Rodrigues formula \cite{whittaker1960}, in the same way as Eq.\ (46) in \cite{littlejohn2010b}. Let us start with vector ${\bf J}_1$. The action of the rotations on this vector can be written

\begin{equation}
\label{ch7: eq_J1_rotation}
R({\bf j}_{13}, 2 \phi_{13}) R(- {\bf j}_6, 2 \phi_6) R({\bf j}_{12}, 2 \phi_{12}) {\bf J}_1 = {\bf J}_1  \, . 
\end{equation}

\begin{figure}[tbhp]
\begin{center}
\includegraphics[width=0.45\textwidth]{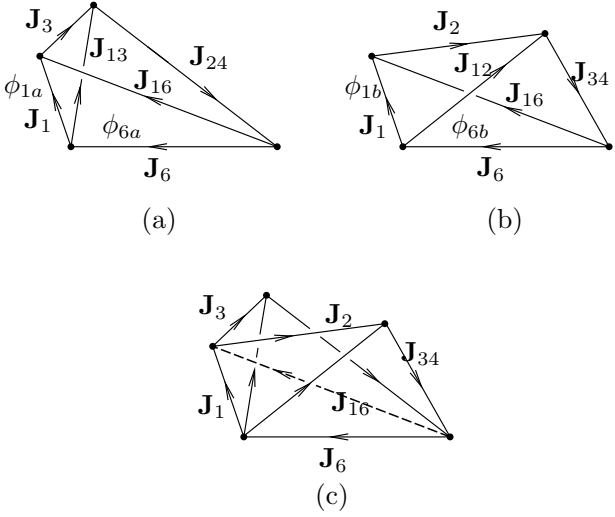}
\caption{Decomposition of the angles $\phi_1$ and $\phi_6$ into dihedral angles in two tetrahedra.}
\label{ch7: fig_angle_split_tetrahedra}
\end{center}
\end{figure}

By inserting an edge ${\bf J}_{16} = {\bf J}_1 + {\bf J}_6$ as in part (c) of Fig.\ \ref{ch7: fig_angle_split_tetrahedra}, we split the angle $\phi_6$ that appears in the middle rotation in Eq.\  (\ref{ch7: eq_J1_rotation})  into two internal dihedral angles $\phi_{6a}$ and $\phi_{6b}$, of the tetrahedrons in Figs.\ \ref{ch7: fig_angle_split_tetrahedra}(a) and \ref{ch7: fig_angle_split_tetrahedra}(b), respectively.

Then the rotations in Eq.\ (\ref{ch7: eq_J1_rotation}) become 

\begin{eqnarray}
&& R({\bf j}_{13}, 2 \phi_{13}) R(- {\bf j}_6, 2 \phi_6) R({\bf j}_{12}, 2 \phi_{12})  \nonumber  \\  \nonumber
&=& [R({\bf j}_{13}, 2 \phi_{13}) R(- {\bf j}_6, 2 \phi_{6a})] [R(- {\bf j}_6, 2 \phi_{6b})  R({\bf j}_{12}, 2 \phi_{12}) ]  \\  \nonumber
&=& R({\bf j}_1, 2 \phi_{1a}) R({\bf j}_1, 2 \phi_{1b})  \\
&=& R({\bf j}_1, 2 \phi_1)  \, , 
\label{ch7: eq_J1_holonomy}
\end{eqnarray}
where we have used the Hamilton-Rodrigues formula twice in the second equality. In the third equality, we used the fact that $\phi_1 = \phi_{1a} + \phi_{1b}$, where the angles $\phi_{1a}$ and $\phi_{1b}$ are internal dihedral angles for the tetrahedra in Figs.\ \ref{ch7: fig_angle_split_tetrahedra}(a) and \ref{ch7: fig_angle_split_tetrahedra}(b), respectively. Thus, we find that the product of the three rotations in Eq.\  (\ref{ch7: eq_J1_rotation}) is $R({\bf j}_1, 2 \phi_1)$, where $\phi_1$ is the angle between the triangle 1-2-12 and the triangle 1-3-13. We can lift the rotation Eq.\  (\ref{ch7: eq_J1_holonomy}) up to $SU(2)$ with the same axis and angle. Its action on the spinor at $p$ is a pure phase. To undo this pure phase, we follow the Hamiltonian flow of $I_1$ by an angle $-2 \phi_1$, modulo $2 \pi$.

\begin{figure}[tbhp]
\begin{center}
\includegraphics[width=0.30\textwidth]{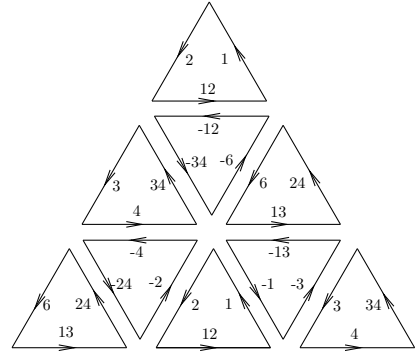}
\caption{The angles $\phi_r$ is the angle between the normals of the adjacent triangles sharing the edge $J_r$, where the normals are defined by the orientation of the triangles shown. This is essentially Fig.\ 2 in \cite{littlejohn2010a}, with an unsymmetrical labeling of the $9j$.}
\label{ch7: fig_outer_normals}
\end{center}
\end{figure}

Similarly, we can find the rotations acting on ${\bf J}_2, {\bf J}_3, {\bf J}_4$, and ${\bf J}_6$, and proceed to calculate the action integral as in \cite{littlejohn2010b}. Instead of completing the derivation of the action integral using our unsymmetrical labeling of the $9j$ symbol, we will quote the result from Eq.\ (12) in \cite{littlejohn2010a}. It is given by 

\begin{equation}
\label{ch7: eq_action_integral}
S^{(1)} =  2 \sum_r \, J_r \psi_r^{(1)}  \, ,   
\end{equation}
where $\psi_r^{(1)} = \pi -  \phi_r$ is the external dihedral angle between the normals of the two triangles adjacent to $J_r$. The orientations of the triangles are defined in Fig.\ \ref{ch7: fig_outer_normals}. The sum is over $r = 1,2,3,4,6, 12, 34, 13, 24$. The relative action integral that corresponds to the other solution $(x_2, y_2)$ of Eq.\  (\ref{ch7: eq_vector_equations}) is 

\begin{equation}
\label{ch7: eq_action_integral_2}
S^{(2)} =  2 \sum_r \, J_r \psi_r^{(2)}  \, , 
\end{equation}
which has the same expression as Eq.\ (\ref{ch7: eq_action_integral}), but we should note that the angles $\psi_r^{(2)}$ are different from $\psi_r^{(1)}$, because the vector configuration has a different set of dot products. As in \cite{littlejohn2010a}, we pick $S^{(1)}$ to correspond to the root in which $- \pi \le \psi_r \le \pi$, and pick $S^{(2)}$ to correspond to the root in which $0 \le \psi_r \le \pi$. 

Altogether, the asymptotic formula for the $9j$ symbol when all $j$'s are large is given by Eq.\ (1) in \cite{littlejohn2010a}, which we reproduce here: 

\begin{eqnarray}
\label{ch7: eq_9j_all_j_large}
&& \left\{
   \begin{array}{ccc}
    j_1 & j_2 & j_{12} \\ 
    j_3 & j_4 & j_{34}   \\ 
    j_{13} & j_{24} & j_5   \\
  \end{array} 
  \right\}    \\   \nonumber
  &=&    \frac{1}{4 \pi  \sqrt{| V_{123}^{(1)} V_{432}^{(1)} - V_{214}^{(1)} V_{341}^{(1)} |}}  \cos (S^{(1)})   \\  \nonumber
  &&    + \frac{1}{ 4 \pi  \sqrt{| V_{123}^{(2)} V_{432}^{(2)} - V_{214}^{(2)} V_{341}^{(2)} |}}  \sin (S^{(2)})   \, .
\end{eqnarray}
It is found from Eq.\ (17) and Eq.\ (18) in  \cite{littlejohn2010a} that, when the configuration goes to its time-reversed image, that is, when all the vectors reverse their directions, the actions transform according to $S^{(1)} \rightarrow - S^{(1)}$ and $S^{(2)} \rightarrow - S^{(2)} + 2 \pi (\sum_{r=1}^9 j_r) + 9 \pi $. As a result, the two terms $\cos(S^{(1)})$ and $\sin(S^{(2)})$ in the $9j$ formula (\ref{ch7: eq_9j_all_j_large}) are invariant under time-reversal symmetry. In the asymptotic formula (\ref{ch7: eq_main_formula_12j}) for the $12j$ symbol that we will derive below, the additional phases generated from the spinor products will break this time-reversal symmetry.

Putting the amplitudes $\Omega$ from Eq.\  (\ref{ch7: eq_amplitude}) and the relative actions $S^{(1)}$  and $S^{(2)}$ into Eq.\  (\ref{ch7: eq_general_formula}), we find 

\begin{widetext}
\begin{eqnarray}
 \braket{b|a}  
&=&  e^{i\kappa_1}  \frac{ \sqrt{J_{12} J_{34} J_{13} J_{24}}}{2 \pi \sqrt{| V_{123}^{(1)} V_{432}^{(1)} - V_{214}^{(1)} V_{341}^{(1)}  |} }    \left[ ( \tau^b(z_{11}) )^\dagger (\tau^a(z_{11}))  +  e^{i ( S^{(1)} - \mu_1 \pi /2) / \hbar}   \left( U_b^{(1)} \tau^b(z_{11}) \right)^\dagger  \left(U_a^{(1)} \tau^a(z_{11})\right)     \right]    
\label{ch7: eq_general_formula_2}    \\   \nonumber
&& +  e^{i\kappa_2}  \frac{ \sqrt{J_{12} J_{23} J_{13} J_{24}}}{2 \pi \sqrt{| V_{123}^{(2)} V_{432}^{(2)} - V_{214}^{(2)} V_{341}^{(2)}  |} }   \left[ ( \tau^b(z_{21}) )^\dagger (\tau^a(z_{21}))  +  e^{i ( S^{(2)} - \mu_2 \pi /2) / \hbar}   \left( U_b^{(2)} \tau^b(z_{21}) \right)^\dagger  \left(U_a^{(2)} \tau^a(z_{21})\right)     \right]
\end{eqnarray}
\end{widetext}
where the superscripts $(1)$ and $(2)$ are labels used to distinguish the first and the second solutions to Eq.\  (\ref{ch7: eq_vector_equations}). Here we have factored out two arbitrary phases $e^{i \kappa_1}$ and $e^{i \kappa_2}$ for the two pairs of stationary phase contributions. The rotation matrices $U_a^{(i)}$, $i = 1,2$, are determined by the paths from $z_{i1}$ to $z_{i2}$ along ${\mathcal L}_a^{9j}$. Similarly the rotation matrices $U_b^{(i)}$, $i = 1,2$, are determined by the paths from $z_{i1}$ to $z_{i2}$ along ${\mathcal L}_b^{9j}$. See Eq.\ (76) in \cite{yu2011} for a similar, but simpler, expression for the case of the $9j$ symbol.

\section{\label{ch7: sec_spinor_products}Spinor Products}

\begin{figure}[tbhp]
\begin{center}
\includegraphics[width=0.30\textwidth]{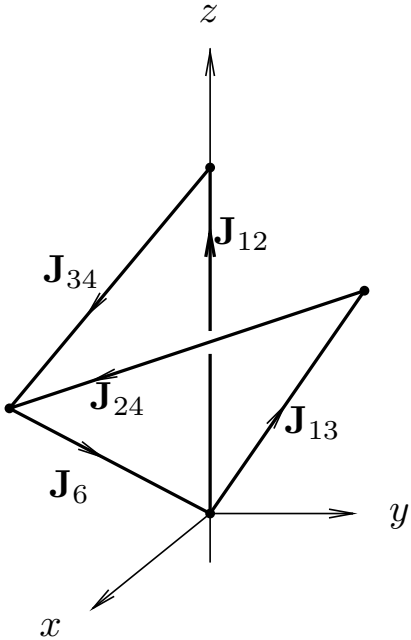}
\caption{The vector configuration at the point $z_{11}$ in $I_{11}$.}
\label{ch7: fig_config_z11}
\end{center}
\end{figure}

We choose the vector configurations associated with $z_{11}$ to correspond to a particular orientation of the vectors. We put ${\bf J}_{12}$ along the $z$ axis, and put ${\bf J}_6$ inside the $xz$ plane, as illustrated in Fig.\ \ref{ch7: fig_config_z11}. Let the inclination and azimuth angles $(\theta, \phi)$ denote the direction of the vector ${\bf J}_{13}$. From Fig.\ \ref{ch7: fig_config_z11}, we see that $\phi$ is the angle between the $({\bf J}_{12}, {\bf J}_6)$ plane and the $({\bf J}_{12}, {\bf J}_{13})$ plane. We denote this angle by $\phi = \phi_{12}$. The inclination angle $\theta$ is the angle between the vectors ${\bf J}_{12}$ and ${\bf J}_{13}$.

The gauge choices for the spinors at the reference point $z_{11}$ are arbitrary, and they only contribute a phase that can be absorbed into $e^{i \kappa_1}$. To be concrete, since ${\bf J}_{12}$ points in the $z$ direction, we choose the spinor $\tau^a(z_{11})$ to be the $\mu^{\text th}$ standard eigenvector for $S_z$, that is,

\begin{equation}
\tau_\alpha^a ( z_{11} ) = \delta_{\alpha \mu}  \, . 
\end{equation}
For the spinor $\tau^b(z_{11})$, we choose it to be an eigenvector of ${\bf J}_{13} \cdot {\bf S}$ in the north standard gauge, that is,

\begin{equation}
\tau_\alpha^b(z_{11}) = e^{i (\alpha - \nu) \phi_{12}} \, d^s_{\nu \alpha } (\theta) \, ,
\end{equation}
where $(\phi_{12}, \theta)$ are the spherical angles of ${\bf J}_{13}$ in a reference frame where ${\bf J}_{12}$ is in the $z$-direction, and ${\bf J}_6$ is in the $xz$-plane. See Fig.\ \ref{ch7: fig_config_z11}. We denote the azimuthal angle by $\phi_{12}$, because it is also the angle at ${\bf J}_{12}$ between the $({\bf J}_{12}, {\bf J}_{13})$ plane and the $({\bf J}_{12}, {\bf J}_{6})$ plane.

Taking the spinor inner product, we obtain  

\begin{equation}
\label{ch7: eq_spinor_prod_11}
(\tau^b(z_{11}))^\dagger (\tau^a(z_{11}))  = e^{- i (\mu - \nu) \phi_{12}} \, d^s_{\nu \mu} (\theta) \, .
\end{equation}
To evaluate the other spinor product at $z_{12}$, we need to find the rotation matrices $U_a^{(1)}$ and $U_b^{(1)}$, which are generated from paths $\gamma_a$ and $\gamma_b$ from $z_{11}$ to $z_{12}$ along ${\mathcal L}_a^{9j}$ and ${\mathcal L}_b^{9j}$, respectively.

We choose the path $\gamma_a$ to be the path from $p$ to $q$ generated by the ${\bf J}_{12}^2$ flow and the ${\bf J}_{34}^2$ flow, which are illustrated in Fig.\ \ref{ch7: fig_loop_large_space} in the large phase space, in Fig.\ \ref{ch7: fig_loop_small_space}(a) in the angular momentum space. This path contains no flow generated by the total angular momentum, so 

\begin{equation}
U_a^{(1)} = 1 \, . 
\end{equation}
We choose the path $\gamma_b$ to be the inverse of the path from $q$ back to $p$ along ${\mathcal L}_b^{9j}$ in Fig.\ \ref{ch7: fig_loop_large_space}, which contains only one overall rotation around $- {\bf j}_6$. Thus

\begin{equation}
U_b^{(1)} = U( {\bf \hat{j}}_6, 2 \phi_6) \, . 
\end{equation}
 
\begin{figure}[tbhp]
\begin{center}
\includegraphics[width=0.30\textwidth]{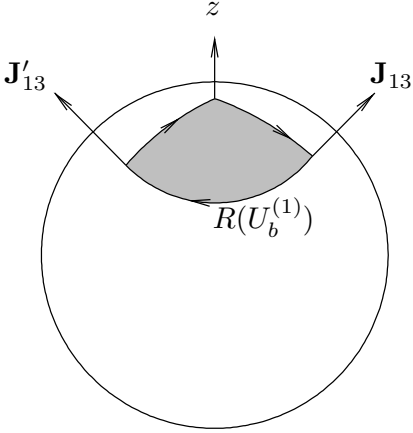}
\caption{The phase difference between two gauge choices can be expressed as an area around a closed loop on the unit sphere.}
\label{ch7: fig_spherical_area}
\end{center}
\end{figure}

The rotation associated with  $U_b^{(1)}$ is illustrated in Fig.\ \ref{ch7: fig_loop_small_space}(b). It effectively moves ${\bf J}_{13}$ to its mirror image ${\bf J}_{13}'$ across the 12-34-6 triangle in the $xz$-plane, which has the direction given by  $(- \phi_{12}, \theta)$. Thus $U_b \, \tau^b(z_{11})$ is an eigenvector of ${\bf J}_{13}' \cdot {\bf S}$, and is up to a phase equal to the eigenvector of ${\bf J}_{13}' \cdot {\bf S}$ in the north standard gauge. Thus, we have

\begin{equation}
[U_b^{(1)} \, \tau^b(z_{11})]_\alpha = e^{i \nu H_{13}}  \, e^{- i (\alpha - \nu) \phi_{12}} \, d^s_{\nu \alpha } (\theta)  \, ,
\end{equation}
where $H_{13}$ is a holonomy phase factor equal to the area of a spherical triangle on a unit sphere; see Fig.\ \ref{ch7: fig_spherical_area}. Therefore, the spinor product at the intersection $I_{12}$ is 

\begin{equation}
\label{ch7: eq_spinor_prod_12}
( U_b^{(1)} \, \tau^b(z_{11}))^\dagger (U_a \tau^a(z_{11})) = e^{i \nu H_{13}}  \, e^{i (\mu - \nu) \phi_{12}} \, d^s_{\nu \mu} (\theta)  \, .
\end{equation}

Let us denote the first term in Eq.\  (\ref{ch7: eq_general_formula_2}) by $T_1$. Substituting the spinor inner products of Eqs.\  (\ref{ch7: eq_spinor_prod_11}) and (\ref{ch7: eq_spinor_prod_12}) into Eq.\  (\ref{ch7: eq_general_formula_2}), we find that $T_1$ is given by

\begin{eqnarray}
\label{ch7: eq_12j_formula_3a}
T_1
&=& \frac{e^{i \kappa_1}   \sqrt{J_{12} J_{34} J_{13} J_{24}}  }{ \pi  \sqrt{ |V_{123} V_{432} - V_{214} V_{341} |}} \, d^s_{\nu \mu} (\theta)    \\  \nonumber
&& \times \cos \left[ S^{(1)}   - \frac{\mu_1 \pi}{4} + \mu \phi_{12} + \nu \left( \frac{H_{13}}{2} - \phi_{12} \right)  \right]  \, . 
\end{eqnarray}

Using a different choice of the reference point and paths, we can derive an alternative expression for the inner product, and eliminate the term $H_{13}$. Let us choose a new reference point $z_{11}$ to correspond to an orientation in which ${\bf J}_{13}$ is along the $z$-axis, and ${\bf J}_6$ lies in the $x$-$z$ plane. We choose the path $\gamma_a$ to go from $p$ to $q'$ along the first two paths in Fig.\ \ref{ch7: fig_loop_large_space}, and we choose $\gamma^b$ to be the inverse of the last two paths that goes from $q'$ back to $p$ in Fig.\ \ref{ch7: fig_loop_large_space}. Through essentially the same arguments, we find 

\begin{eqnarray}
 \label{ch7: eq_12j_formula_3b}
T_1
&=& \frac{e^{i \kappa_1}   \sqrt{J_{12} J_{34} J_{13} J_{24}}   }{\pi  \sqrt{ |V_{123} V_{432} - V_{214} V_{341} |}} \, d^s_{\nu \mu} (\theta)   \,   \\  \nonumber
&& \times \cos \left[ S^{(1)}  - \frac{\mu_1 \pi}{4} +  \mu \left(\frac{H_{12}}{2} - \phi_{13} \right)  +  \nu \phi_{13}  \right]  \, . 
\end{eqnarray}
Here $H_{12}$ is another holonomy for the ${\bf J}_{12}$ vector, and the angle $\phi_{13}$ is the angle between the (${\bf J}_{13}$, ${\bf J}_{12}$) plane and (${\bf J}_{13}$, ${\bf J}_6$) plane.  Because the quantities $\psi_i,  \phi_{12}, \phi_{13}, H_{12}, H_{13}$ depend only on the geometry of the vector configuration, and are independent of $\mu$ and $\nu$, we conclude that the argument in the cosine must be linear in $\mu$ and $\nu$. Equating the two arguments of the cosine in Eqs.\  (\ref{ch7: eq_12j_formula_3a}) and (\ref{ch7: eq_12j_formula_3b}), we find that this linear term is $( \mu \phi_{12} + \nu \phi_{13} )$. Using the Maslov index $\mu_1 = 0$ from \cite{littlejohn2010a}, we find

\begin{eqnarray}
\label{ch7: eq_main_formula_wo_phase1}
T_1  
&=& \frac{e^{i \kappa_1}   \sqrt{J_{12} J_{34} J_{13} J_{24}}   }{\pi \sqrt{ |  V_{123}^{(1)} V_{432}^{(1)} - V_{214}^{(1)} V_{341}^{(1)}  |}}  \,  d^s_{\nu \mu} (\theta^{(1)})  \\  \nonumber
&&    \, 
  \times \cos \left( S^{(1)} + \mu \phi_{12}^{(1)} + \nu \phi_{13}^{(1)}  \right)  \, ,
\end{eqnarray}
where we have put back the superscript $(1)$. Through an analogous calculation, we find

\begin{eqnarray}
\label{ch7: eq_main_formula_wo_phase2}
T_2
&=& \frac{e^{i \kappa_2 } \, \sqrt{J_{12} J_{34} J_{13} J_{24}}   }{ \pi \sqrt{ | V_{123}^{(2)} V_{432}^{(2)} - V_{214}^{(2)} V_{341}^{(2)}  |}}  \,  d^s_{\nu \mu} (\theta^{(2)}) \,   \\   \nonumber
&&   \, \times  \sin \left( S^{(2)} +  \mu \phi_{12}^{(2)} + \nu \phi_{13}^{(2)} \right)  \, .
\end{eqnarray}  \\

\section{\label{ch7: sec_12j_formula}Asymptotic Formula for the $12j$ Symbol}

From the definition in Eq.\  (\ref{ch7: eq_12j_definition}), we see that the factor $( [j_{12}][j_{34}] [j_{13}][j_{24}] )^{1/2}$ in the denominator of Eq.\  (\ref{ch7: eq_12j_definition}) partially cancels out the factor $( J_{12} J_{34} J_{13} J_{24} )^{1/2}$ from $T_1$ and $T_2$ in Eqs.\  (\ref{ch7: eq_main_formula_wo_phase1}) and (\ref{ch7: eq_main_formula_wo_phase2}), respectively, leaving a constant factor of $1/4$. Because the $12j$ symbol is a real number, the relative phase between $e^{i \kappa_1}$ and $e^{i \kappa_2}$ must be $\pm 1$. Through numerical experimentation, we found it to be $+1$. We use the limiting case of $j_5 = s = 0$ from Eq.\ (A9) in \cite{jahn1954} to determine the overall phase convention. This determines most of the overall phase. The rest can be fixed through numerical experimentation. Putting the pieces together, we obtain a new asymptotic formula for the $12j$ symbol with one small quantum number:

\begin{widetext}
\begin{eqnarray}
\label{ch7: eq_main_formula_12j}
 \left\{
  \begin{array}{cccc}
    j_1 & j_2 & j_{12} & j_{125} \\ 
    j_3 & j_4 & j_{34} &  j_{135}  \\ 
    j_{13} & j_{24} & s & j_6  \\
  \end{array} 
  \right\}   
=  \frac{(-1)^{\mu}}{ 4 \pi \, \sqrt{ (2j_{125}+1)(2j_{135}+1) }} \, 
&&  \left[  \frac{d^{s}_{\nu \, \mu} (\theta^{(1)})}{ \sqrt{| V_{123}^{(1)} V_{432}^{(1)} - V_{214}^{(1)} V_{341}^{(1)} | } } \cos (S^{(1)}  + \mu \phi_{12}^{(1)} + \nu \phi_{13}^{(1)})  \right.  \\  \nonumber
&&  \left.  \quad  \,   + \frac{d^{s}_{\nu \, \mu} (\theta^{(2)})}{ \sqrt{| V_{123}^{(2)} V_{432}^{(2)} - V_{214}^{(2)} V_{341}^{(2)} |}}  \sin (S^{(2)}  + \mu \phi_{12}^{(2)} + \nu \phi_{13}^{(2)}) \right]  \, .
\end{eqnarray}
\end{widetext}
As mentioned above, the additional terms from the spinor product break the time-reversal symmetry. Thus, it is essential that $S^{(1)}$ and  $S^{(2)}$ are evaluated at the configurations in which $V = {\mathbf J}_6 \cdot ({\mathbf J}_{12} \times {\mathbf J}_{13})  < 0$, and not at their mirror images.

Here, the indices on the $d$-matrix are given by $\mu = j_{125}-j_{12}$ and $\nu = j_{135}-j_{13}$. They are of the same order as the small parameter $s$. The phases $S^{(1)}$ and $S^{(2)}$ are defined in (\ref{ch7: eq_action_integral}), and the $V$'s are defined by

\begin{equation}
V_{ijk} = {\bf J}_i \cdot ({\bf J}_j \times {\bf J}_k) \, . 
\end{equation}

\begin{figure}[tbhp]
\begin{center}
\includegraphics[width=0.28\textwidth]{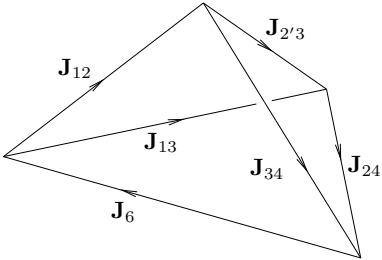}
\caption{The angles $\phi_{12}$ and $\phi_{13}$ are internal dihedral angles in the tetrahedron with the six lengths  ${\bf J}_6, {\bf J}_{12}, {\bf J}_{34}, {\bf J}_{13}, {\bf J}_{24}, {\bf J}_{2'3}$,  where ${\bf J}_{2'3} = {\bf J}_3 - {\bf J}_2$. The angle $\theta$ is the angle between ${\bf J}_{12}$ and ${\bf J}_{13}$. }
\label{ch7: fig_tetrahedron_9j}
\end{center}
\end{figure}

The angles $\phi_{12}$ and $\phi_{13}$ are internal dihedral angles at the edge $J_{12}$ and $J_{13}$, respectively, of a tetrahedron formed by the six vectors ${\bf J}_{12}, {\bf J}_{13}, {\bf J}_{24}, {\bf J}_{34}, {\bf J}_6$, and ${\bf J}_{2'3}$, where ${\bf J}_{2'3} = {\bf J}_3 - {\bf J}_2$. This tetrahedron is illustrated in Fig.\ \ref{ch7: fig_tetrahedron_9j}. The angle $\theta$ is the angle between the vectors ${\bf J}_{12}$ and ${\bf J}_{13}$. The explicit expression for the angles $\phi_{12}$, $\phi_{13}$, and $\theta$ are given by the following equations

\begin{eqnarray}
\label{ch7: eq_phi_12_def}
\phi_{12} &=&  \pi - \cos^{-1} \left( \frac{ ({\bf J}_{12} \times {\bf J}_{13} ) \cdot ({\bf J}_{12} \times {\bf J}_{6} ) }{ | {\bf J}_{12} \times {\bf J}_{13} | \,  | {\bf J}_{12} \times {\bf J}_{6}  |} \right) \, ,    \\
\label{ch7: eq_phi_13_def}
\phi_{13} &=& \pi -  \cos^{-1} \left(  \frac{ ({\bf J}_{13} \times {\bf J}_{12} ) \cdot ({\bf J}_{13} \times {\bf J}_{6} ) }{ | {\bf J}_{13} \times {\bf J}_{12} | \,  | {\bf J}_{13} \times {\bf J}_{6}  |}  \right)  \, ,   \\
\label{ch7: eq_theta_def}
\theta &=& \cos^{-1} \left( \frac{ {\bf J}_{12} \cdot {\bf J}_{13} }{J_{12} J_{13} }  \right)  \, . 
\end{eqnarray}

\section{\label{ch7: sec_plots}Plots}

\begin{figure}[tbhp]
\begin{center}
\includegraphics[width=0.50\textwidth]{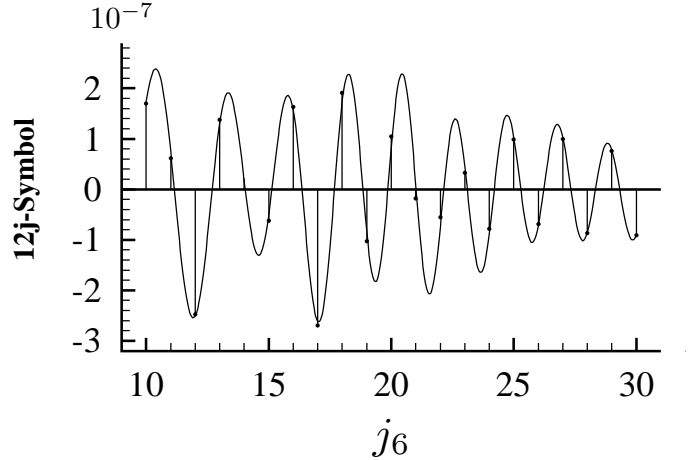}
\caption{Comparison of the exact $12j$ symbol (vertical sticks and dots) and the asymptotic formula (\ref{ch7: eq_main_formula_12j}) in the classically allowed region away from the caustics, for the values of $j$'s shown in Eq.\  (\ref{ch7: eq_12j_values_1_11_case1}). }
\label{ch7: fig_12j_plot_1_case1}
\end{center}
\end{figure}

\begin{figure}[tbhp]
\begin{center}
\includegraphics[width=0.50\textwidth]{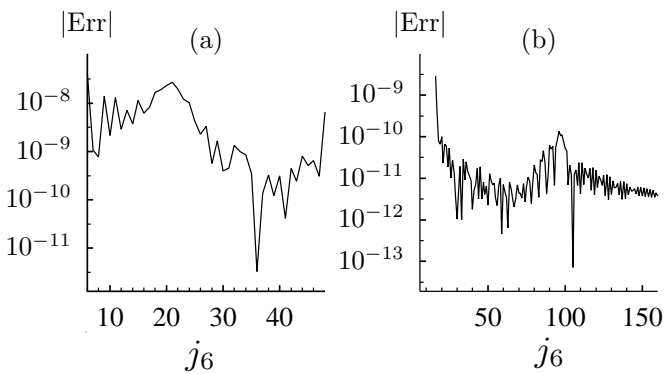}
\caption{Absolute value of the error of the asymptotic formula (\ref{ch7: eq_main_formula_12j}) for (a) the case shown in Eq.\  (\ref{ch7: eq_12j_values_1_11_case1}), and (b) the case shown in Eq.\  (\ref{ch7: eq_12j_values_1_11}). The error is defined as the difference between the approximate value and the exact value. }
\label{ch7: fig_12j_11_1_errors}
\end{center}
\end{figure}

We illustrate the accuracy of the approximation Eq.\  (\ref{ch7: eq_main_formula_12j}) by plotting it against the exact $12j$ symbol in the classically allowed region for the following values of the $j$'s:  

\begin{equation}
\label{ch7: eq_12j_values_1_11_case1}
\left\{
  \begin{array}{cccc}
    j_1 & j_2 & j_{12} & j_{125} \\ 
    j_3 & j_4 & j_{34} &  j_{135}  \\ 
    j_{13} & j_{24} & s_5 & j_6  \\
  \end{array} 
  \right\}   
=
	\left\{
  \begin{array}{rrrr}
    51/2 & 59/2 & 21 & 22 \\ 
    55/2 & 53/2 & 27 & 26 \\ 
    27 & 25 & 1 & j_6 \\
  \end{array} 
  \right\} \, . 
\end{equation}
The result is shown in Fig.\ \ref{ch7: fig_12j_plot_1_case1}. From the error plot in Fig.\ \ref{ch7: fig_12j_11_1_errors}(a), we see that the agreement is excellent, even for these relatively small values of the $j$'s.

\begin{figure}[tbhp]
\begin{center}
\includegraphics[width=0.50\textwidth]{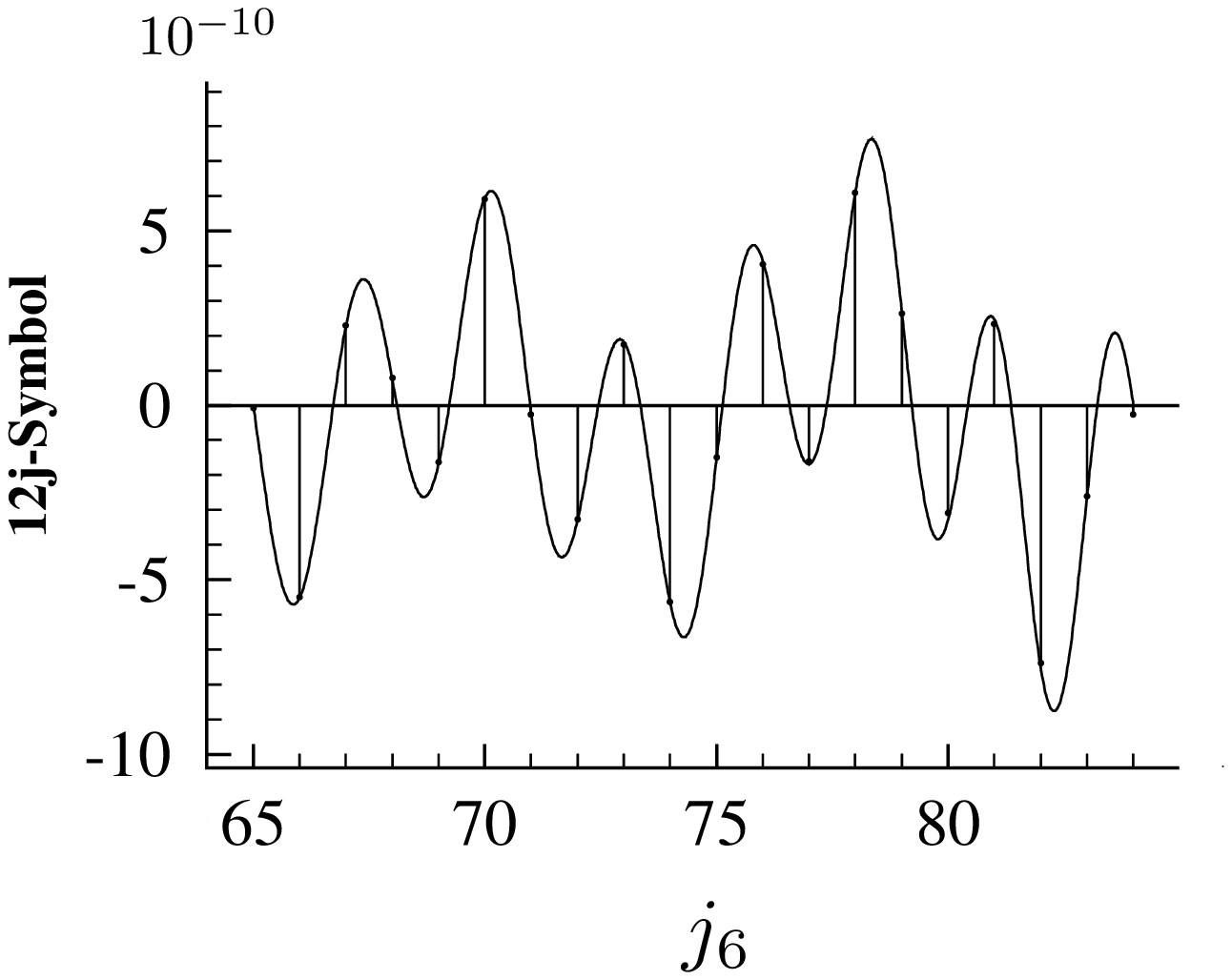}
\caption{Comparison of the exact $12j$ symbol (vertical sticks and dots) and the asymptotic formula (\ref{ch7: eq_main_formula_12j}) in the classically allowed region away from the caustics, for the values of $j$'s shown in Eq.\  (\ref{ch7: eq_12j_values_1_11}). }
\label{ch7: fig_12j_plot_1}
\end{center}
\end{figure}

Since the asymptotic formula (\ref{ch7: eq_main_formula_12j}) should become more accurate as the values of the $j$'s get larger, we plot the formula against the exact $12j$ symbol for another example,  

\begin{equation}
\label{ch7: eq_12j_values_1_11}
\left\{
  \begin{array}{cccc}
    j_1 & j_2 & j_{12} & j_{125} \\ 
    j_3 & j_4 & j_{34} &  j_{135}  \\ 
    j_{13} & j_{24} & s_5 & j_6  \\
  \end{array} 
  \right\}   
=
	\left\{
  \begin{array}{rrrr}
    211/2 & 219/2 & 91 & 92 \\ 
    205/2 & 223/2 & 107 & 108 \\ 
    99 & 93 & 2 & j_6 \\
  \end{array} 
  \right\} \, ,
\end{equation}
in the classically allowed region away from the caustic in Fig.\ \ref{ch7: fig_12j_plot_1}. These values of the $j$'s are roughly four times those in Eq.\  (\ref{ch7: eq_12j_values_1_11_case1}). The errors for this case are displayed in Fig.\ \ref{ch7: fig_12j_11_1_errors}(b). By comparing Figs.\ \ref{ch7: fig_12j_11_1_errors}(a) and \ref{ch7: fig_12j_11_1_errors}(b), we can conclude that the error scales with the $j$'s.

\section{Conclusions}

In this paper, we have derived an asymptotic formula of the $12j$ symbol with one small angular momentum, generalizing the special formula of the $12j$ symbol, Eq.\ (A9) in \cite{jahn1954}. By looking at the other special formula for the $12j$ symbol, Eq.\ (A8) in \cite{jahn1954}, we can guess that the other asymptotic limit of the $12j$ symbol will involve the semiclassical analysis of the trivial $9j$ symbol, which reduces to a product of two $6j$ symbols. We will present that result in a future paper.

The analysis of the $12j$ symbol in this paper is a natural extension of the analysis of the $9j$ symbol in \cite{yu2011}. Based on the calculations in these two papers, we can summarize our steps in finding asymptotic formulas for the $3nj$ symbols with small and large quantum numbers. First, we ignore the small quantum numbers and any of the large quantum numbers that involve the indices of the small ones. For instance, in this paper, $j_5 = s_5$ is small, so we ignore $j_5$, $j_{125}$, and $j_{135}$. The remaining relevant large quantum numbers determine the Lagrangian manifolds. Once we fix the Lagrangian manifolds, the scalar WKB parts of the wave-functions can be derived from a semiclassical analysis of these Lagrangian manifolds, following the procedure in \cite{littlejohn2007, littlejohn2010b}. The spinor parts of the wave-functions at the intersection points of the Lagrangian manifolds are determined by the path used to calculate the action integral in the semiclassical analysis. Finally, taking the inner product of both the scalar part and the spinor part of the wave-functions, we can derive an asymptotic formula for the $3nj$ symbol with small and large angular momenta. 

In general, we note that the asymptotic limits of a $3nj$ symbol with one small angular momentum is expressed in terms of the geometry associated with the asymptotic limits of a $3mj$-symbol, where $m=n-1$. Since the Wigner $15j$ symbol is used extensively in loop quantum gravity and topological quantum field theory, we suspect that there are deeper, and more geometrical interpretations of these approximate relations of the $3nj$ symbol in their various semiclassical limits.


\bibliography{SLQN12j}

\end{document}